\documentclass[useAMS,usenatbib]{mnras}
\usepackage[T1]{fontenc}
\usepackage{amsmath, amssymb, bm}
\usepackage{xcolor}
\usepackage{graphicx}
\usepackage{caption}
\usepackage{subcaption}
\usepackage{booktabs}
\usepackage{tabularx}
\usepackage{catchfile}
\usepackage{hyperref}
\usepackage[capitalize]{cleveref}
\usepackage{acro}
\usepackage{orcidlink}

\hypersetup{breaklinks=true}

\newcommand{\zCMB}{z_{\rm CMB}}
\newcommand{\mlim}{m_{\rm lim}}

\newcommand{\Vext}{\bm{V}_{\rm ext}}

\newcommand{\Mpch}{\ensuremath{\,h^{-1}\,\mathrm{Mpc}}}
\newcommand{\Mpc}{\ensuremath{\,\mathrm{Mpc}}}
\newcommand{\kmsec}{\ensuremath{\,\mathrm{km}\,\mathrm{s}^{-1}}}
\newcommand{\kmsecMpc}{\ensuremath{\,\mathrm{km}\,\mathrm{s}^{-1}\,\mathrm{Mpc}^{-1}}}

\newcommand{\mobs}{m_{\rm obs}}
\newcommand{\cobs}{c_{\rm obs}}

\newcommand{\dd}{{\rm d}}
\newcommand{\Om}{\Omega_{\rm m}}
\newcommand{\logZ}{\log \mathcal{Z}}
\newcommand{\TWOMPP}{2M\texttt{++}}
\newcommand{\Manticore}{\texttt{Manticore-Local}}

\DeclareAcronym{CMB}{short = CMB, long  = cosmic microwave background}
\DeclareAcronym{LOS}{short = LOS, long  = line-of-sight}
\DeclareAcronym{BORG}{short = \texttt{BORG}, long  = \textit{Bayesian Origin Reconstruction from Galaxies}}
\DeclareAcronym{LCDM}{short = $\Lambda$CDM, long  = $\Lambda$-cold dark matter}
\DeclareAcronym{LMC}{short = LMC, long = Large Magellanic Cloud}
\DeclareAcronym{HST}{short = HST, long = Hubble Space Telescope}
\DeclareAcronym{JWST}{short = JWST, long = James Webb Space Telescope}
\DeclareAcronym{TRGB}{short = TRGB, long = tip of the red giant branch}
\DeclareAcronym{SN}{short = SN, long  = supernova, short-plural = e, long-plural  = e}
\DeclareAcronym{SHOES}{short = SH0ES, long  = Supernovae and $H_0$ for the Equation of State of dark energy}
\DeclareAcronym{CCHP}{short = CCHP, long  = Carnegie--Chicago Hubble Program}
\DeclareAcronym{EDD}{short = EDD, long  = Extragalactic Distance Database}
\DeclareAcronym{KS}{short = KS, long = Kolmogorov--Smirnov}
\DeclareAcronym{PPD}{short = PPD, long = posterior predictive distribution}

\title[$H_0$ from TRGB alone]{Two-rung ladder: $H_0$ from Tip of the Red Giant Branch and geometric anchors alone}

\author[R. Stiskalek et al.]{Richard Stiskalek$^{1}$\thanks{\href{mailto:richard.stiskalek@physics.ox.ac.uk}{richard.stiskalek@physics.ox.ac.uk}}\orcidlink{0000-0002-0986-314X},
Harry Desmond$^{2}$\orcidlink{0000-0003-0685-9791}
and Guilhem Lavaux$^{3}$\orcidlink{0000-0003-0143-8891}
\\
$^{1}$Astrophysics, University of Oxford, Denys Wilkinson Building, Keble Road, Oxford, OX1 3RH, UK\\
$^{2}$Institute of Cosmology \& Gravitation, University of Portsmouth, Dennis Sciama Building, Portsmouth, PO1 3FX, UK\\
$^{3}$CNRS \& Sorbonne Universit\'e, Institut d'Astrophysique de Paris (IAP), UMR 7095, 98 bis bd Arago, F-75014 Paris, France
\\
}

\date{Accepted XXX. Received YYY; in original form ZZZ}
\pubyear{2026}

\begin{document}\label{firstpage}
\pagerange{\pageref{firstpage}--\pageref{lastpage}}
\maketitle

\begin{abstract}
We infer the Hubble constant $H_0$ from the ${\sim}400$ Extragalactic Distance Database hosts with a tip of the red giant branch (TRGB) distance, calibrated on the geometric anchors of the Large Magellanic Cloud and NGC\,4258.
We forward-model the tip magnitudes and host redshifts including magnitude-limited selection, sky exposure, and density-dependent galaxy bias, conditioning on each of the $80$ \Manticore\ realisations of the local density and peculiar-velocity fields in turn.
As most hosts lie at $c\zCMB\lesssim1000\kmsec$, where peculiar velocities are important, the inference is limited by the variance of these fields rather than by the TRGB calibration.
For a fixed realisation, the redshift likelihood and density smoothing each move $H_0$ by less than $1\kmsecMpc$ and a velocity monopole by $5.4\pm3.4\kmsecMpc$, whereas the field-to-field scatter in $H_0$ is $5.2\kmsecMpc$.
Weighting the posteriors from each realisation equally gives $H_0=68.4\pm5.9\kmsecMpc$, while weighting each by its Bayesian evidence gives $72.3\pm2.6\kmsecMpc$.
The latter value is set by a single particularly high-evidence realisation, making it sensitive to the finite sampling of the reconstruction posterior.
This ladder therefore cannot yet arbitrate the Hubble tension.
Extending the TRGB data beyond the present ${\sim}1000\kmsec$, together with improved reconstructions of the local Universe, would reduce the field-to-field scatter well below the $5.2\kmsecMpc$ we find here.
\end{abstract}


\section{Introduction}\label{sec:intro}

The two high-precision local measurements of the Hubble constant $H_0$ both employ a Type~Ia \ac{SN} rung: \ac{SHOES} anchors on Cepheid--\ac{SN} hosts~\citep{Riess_2016,Riess_2018,Riess_2021,Riess_2022,Breuval_2024}, and the \ac{CCHP} on \ac{TRGB}--\ac{SN} hosts~\citep{Beaton_2016,Freedman_2019,Freedman_2025,Freedman_2025_erratum}.
Although the two agree at the ${\sim}1.2\,\sigma$ level, \ac{SHOES} sits in ${\sim}5\,\sigma$ tension with \textit{Planck} \ac{CMB} constraints under \acs{LCDM}, whereas the \ac{CCHP} sits in ${\sim}1.6\,\sigma$ tension with the same reference~\citep{Planck_2020_cosmo,Tristram_2024,Freedman_2025,Freedman_2025_erratum}.
The \ac{SN} rung itself carries standardisation, host-selection, covariance, dust, and absolute-magnitude calibration systematics that propagate into the inferred $H_0$~\citep[e.g.][]{Mandel_2022,Efs_SNe,Wojtak_2025,Carreres_2025}, motivating local-ladder tests that do not use it.

An alternative route is to infer $H_0$ directly from galaxies with both \ac{SN}-independent distances and measured redshifts, omitting the \ac{SN} rung entirely~\citep{Hubble,Willick_2001,Kenworthy_2022}.
Exemplifying this approach, \citet{Stiskalek_2026} forward-model the apparent magnitudes of individual Cepheid stars and the redshifts of their host galaxies in a hierarchical Bayesian model, and describe the local peculiar-velocity field with the \Manticore\ reconstruction~\citep{McAlpine_2025}.
They obtain $H_0=71.1\pm1.4\kmsecMpc$~\citep{Stiskalek_2026,Stiskalek_2026_erratum}.
Other routes that drop the Type~Ia rung replace it with surface-brightness fluctuations~\citep{Blakeslee_2021,Cantiello_2023,Jensen_2025}, Tully--Fisher catalogues~\citep{Schombert,Tully_2023,Boubel_2024B}, or Type~II \acp{SN}~\citep{deJaeger_2022,Vogl_2024}.
These routes yield $H_0$ values typically in the range ${\sim}73$--$76\kmsecMpc$.

Water masers stand apart from these routes, providing a geometric, calibration-independent $H_0$ measurement that is not tied to any stellar standard candle~\citep{Lo_2005}.
The sub-parsec disc of NGC\,4258 was resolved by~\citet{Miyoshi_1995} and its geometric distance measured by~\citet{Herrnstein_1999}.
The method was later extended to a multi-galaxy sample by the Megamaser Cosmology Project (e.g.~\citealt{Braatz_2007,Reid_2009,megamaser}) and recently revisited by~\citet{Stiskalek_2026C}.
Gravitational-wave standard sirens are similarly calibration-independent, with the luminosity distance set by the waveform amplitude.
The redshift is supplied by an electromagnetic counterpart (GW170817;~\citealt{Abbott_2017}) or probabilistically by a galaxy catalogue when the event has no such counterpart (dark sirens; e.g.~\citealt{Abbott_2023}).

The \ac{TRGB} is the sharp cut-off at the bright end of the luminosity function of red giants, produced when their electron-degenerate helium cores ignite at a nearly fixed mass and therefore, at fixed composition, at a nearly fixed bolometric luminosity~\citep{Salaris_1997,Serenelli_2017}.
\citet{Mould_1986} first used the tip to measure a distance, and~\citet{DaCosta_1990} calibrated the giant branch as a standard sequence, but the measurement was only automated by~\citet{Lee_1993}, who located the tip with a Sobel edge-detection filter applied to the luminosity function.
\citet{Madore_1995} then simulated the conditions under which that measurement is unbiased, showing that it degrades at low photometric signal-to-noise and with crowding and contamination by populations other than red giants, and requires ${\gtrsim}100$ stars within one magnitude of the tip.
The \ac{TRGB} is a Population~II halo indicator with different dust, star-formation, and metallicity systematics from Cepheids~\citep{Anand_2022,Bhardwaj_2023,Li_2024b}.

In this work we construct a distance ladder from \ac{TRGB} hosts and geometric anchors.
The closest work in the literature is that of~\citet{Kim_2020}, who infer $H_0=65.8\pm3.5\,({\rm stat})\pm2.4\,({\rm sys})\kmsecMpc$ from \ac{TRGB} distances and redshifts of $33$ galaxies between the Local Group and Virgo, modelling their spherical infall towards Virgo.
Their spherical infall treatment follows the velocity--distance relation of~\citet{Peirani_2006}, who apply it separately to the Local Group and to Virgo.
The same single-structure approach has since been extended to the Coma cluster~\citep{Benisty_2026}.
Here we perform the direct \ac{TRGB} analogue of the Cepheid-only analysis: an $H_0$ inference from a \ac{TRGB} catalogue and the host redshifts alone.

The \ac{EDD} \ac{TRGB} catalogue of~\citet{Anand_2021} is well suited to this exercise.
It provides nearly $500$ hosts, about an order of magnitude more than the \ac{SHOES} Cepheid-host sample.
The sample is nevertheless local, with all hosts at \ac{CMB}-frame recession velocities ${\lesssim}2150\kmsec$ and the bulk at ${\lesssim}1000\kmsec$.
Peculiar velocities are then comparable to the cosmological signal and must be carefully modelled~\citep{Kaiser_1988,Hui_2006,Davis_2011,Carrick_2015,McAlpine_2025,VF_olympics}.
It is also assembled from heterogeneous \ac{HST} programmes with no single well-defined selection function, which can be a significant source of systematic uncertainty~\citep{Hinton_2017,Desmond_2025,Stiskalek_2026}.
A \ac{TRGB}-only inference therefore tests a different combination of stellar calibration and sample selection from the Cepheid-only inference, over a smaller redshift range (${\lesssim}1000\kmsec$ versus $3300\kmsec$).

We forward-model the tip magnitudes, the host redshifts, and the catalogue selection, with the \ac{TRGB} anchored to the geometric distances of the \acl{LMC} (\ac{LMC};~\citealt{Pietrzynski_2019}) and NGC\,4258~\citep{Reid_2019}.
In this regime the peculiar-velocity reconstruction is the dominant modelling systematic.
We therefore use \Manticore~\citep{McAlpine_2025}, a field-level density and velocity reconstruction of the local Universe built with the \ac{BORG} algorithm~\citep{Jasche_2013,Lavaux_2016,Jasche_2019}.
The reconstruction is constrained by \TWOMPP, an all-sky redshift compilation out to ${\sim}200\Mpch$~\citep{Lavaux_2011}.
Field-level \ac{BORG} reconstructions of this type were recently benchmarked as the most accurate of the available low-redshift velocity-field models~\citep{VF_olympics}.
\Manticore\ itself attains the highest Bayesian evidence across independent peculiar-velocity catalogues~\citep{McAlpine_2025}.
We find that the $5.2\kmsecMpc$ scatter in $H_0$ across the \Manticore\ realisations, rather than the \ac{TRGB} calibration, limits the precision of this ladder.

The structure of this paper is as follows.
In~\cref{sec:data} we describe the \ac{TRGB} sample and the geometric anchors.
In~\cref{sec:methods} we set out the hierarchical forward model and the velocity-reconstruction variants.
In~\cref{sec:results} we present the results.
In~\cref{sec:discussion} we interpret them, compare with previous work, and set out the caveats.
In~\cref{sec:conclusions} we summarise our findings and conclude.

All logarithms are base-$10$ unless otherwise stated.
We use $\mathcal{N}(x;\,\mu,\,\sigma^2)$ to denote a one-dimensional normal distribution with mean $\mu$ and variance $\sigma^2$ evaluated at $x$; in higher dimensions $\mu$ is a vector and $\sigma^2$ is replaced by a covariance matrix.
A variable $x$ drawn from this distribution is denoted $x\hookleftarrow\mathcal{N}(\mu,\,\sigma^2)$.
We follow the convention that $h\equiv H_0/(100\kmsecMpc)$.

\section{TRGB hosts and anchors}\label{sec:data}

We use the \ac{TRGB} catalogue of~\citet{Anand_2021}, who provide measurements from \ac{HST} colour--magnitude diagrams of $556$ nearby galaxies, of which $489$ have a tip measurement.
Here and below, F606W and F814W denote the \ac{HST} filter passbands used for the tip colours and magnitudes.
The catalogue distances adopt the F814W colour-standardisation calibration of~\citet{Rizzi_2007} used in the \ac{EDD} reductions~\citep{Anand_2021,Anand_2022}.
Our \ac{EDD} retrieval, made on 2026 March 12, returns $487$ entries, two fewer than the $489$ tip measurements of~\citet{Anand_2021}.
The difference arises because the retrieval joins their catalogue to the Cosmicflows-4~\citep{Tully_2023} \ac{TRGB} distance table and retains only hosts present in both.
From these we remove NGC\,4258 and its satellite NGC\,4258-DF6, since NGC\,4258 enters the model as a geometric anchor with the F814W tip magnitude of~\citet{Jang_2021}.
We further drop six galaxies without a finite F814W tip magnitude, one without the \ac{EDD} colour-standardisation term, and $33$ without a tabulated redshift.
This leaves $445$ hosts with extinction-corrected tip magnitudes $\mobs\in[17.93,\,27.65]$ mag.

The baseline inference applies a bright-end truncation $\mobs>22.1$ mag, removing $44$ nearby objects with tip distances ${\lesssim}\,2\Mpc$ and leaving $401$ hosts.
The \ac{CMB}-frame redshifts of the removed objects are typically negative and hence completely dominated by the peculiar velocity.
The removed objects are predominantly Local Group members, with more than half ($26$ of $44$) belonging to the M31 subgroup.
The Galactic-plane mask of the selection model ($|b|\geq10^\circ$; \cref{sec:angular_selection}) removes a further $7$ hosts, so $394$ hosts enter the likelihood.
We treat the removal of the hosts lacking a colour-standardisation term or a tabulated redshift as a data-availability cut rather than as a selection effect requiring explicit modelling.
The bright-end magnitude truncation and the Galactic-plane mask are instead modelled explicitly in the inference.
The redshift-availability cut is not random in distance: the $33$ hosts without a tabulated redshift are dwarf companions of the M81 and Centaurus\,A groups at $3$--$10\Mpc$.
Whether a redshift has been measured depends on host luminosity rather than on the tip magnitude or colour at fixed distance, so the removal should not bias the absolute calibration $M_0$.
It does deplete the sample towards two nearby groups over that distance range, which the magnitude and angular selection terms do not describe.
We do not test this aspect.
We convert the catalogue redshift of each host to the \ac{CMB} rest frame, $\zCMB$, and take the host sky position from the catalogue.
We use the individual host redshift throughout and do not substitute a group-averaged value.

For each host the \ac{EDD} sample provides the observed F814W tip magnitude $m_{\rm F814W}$ with its $1\sigma$ bounds $(m_{\rm F814W}^{\rm lo},\,m_{\rm F814W}^{\rm hi})$, the F606W--F814W colour with analogous bounds, the F814W extinction correction $A_{814}$, and the colour-standardised absolute magnitude $M_{\rm TRGB}^{\rm EDD}$ used to compute the tabulated \ac{EDD} distance modulus.
We set the extinction-corrected tip magnitude as
\begin{equation}\label{eq:mobs}
    \mobs \equiv m_{\rm F814W} - A_{814},
\end{equation}
and we define the colour entering the \ac{TRGB} standardisation as
\begin{equation}\label{eq:cobs}
    \cobs \equiv c_\star^{\rm EDD} + \frac{M_{\rm TRGB}^{\rm EDD} + 4.06}{0.20},
\end{equation}
with $c_\star^{\rm EDD}=1.23$ mag.
This recovers the published \ac{EDD} colour standardisation exactly.
We invert the published standardisation because the inference takes the extinction-corrected tip magnitude and colour as its observables.
We set the statistical uncertainty on $\mobs$ to $e_m=(m_{\rm F814W}^{\rm hi}-m_{\rm F814W}^{\rm lo})/2$.
We take $e_c$ from the raw F606W--F814W colour bounds when available, and set $e_c=0$ when the colour-standardisation term is supplied only through $M_{\rm TRGB}^{\rm EDD}$.
The halving convention is not important because $e_m$ enters in quadrature with the intrinsic \ac{TRGB} scatter of~\cref{eq:m_obs_lik}, which is typically twice as large.
The catalogue does not report an uncertainty on $A_{814}$, so we treat it as a fixed correction and do not propagate it into $e_m$.
Across the retained sample, $e_m$ has median $0.050$ mag with 16\textsuperscript{th}--84\textsuperscript{th} percentile range $[0.020,\,0.115]$ mag, and $e_c$ has median $0.025$ mag with range $[0.005,\,0.070]$ mag.
By neglecting the $A_{814}$ uncertainty, we implicitly assume it to be subdominant to the statistical uncertainty of $m_{\rm F814W}$.
Any additional scatter can also be absorbed into the intrinsic \ac{TRGB} scatter $\sigma_{\rm int}$.

\begin{figure*}
    \centering
    \begin{subfigure}[t]{0.49\textwidth}
        \centering
        \includegraphics[width=\linewidth]{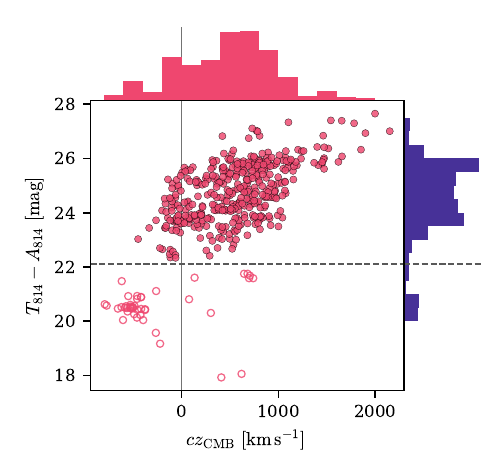}
        \caption{Extinction-corrected F814W tip magnitude as a function of \ac{CMB}-frame recession velocity.
        Marginal histograms show the corresponding one-dimensional velocity and magnitude distributions.
        The dashed line marks the bright-end truncation $\mobs>22.1$ mag.
        Filled points are the $401$ retained hosts and open circles the $44$ removed hosts.
        The removed hosts lie predominantly at negative \ac{CMB}-frame recession velocities, where the peculiar velocity dominates.}
        \label{fig:trgb_magnitude_redshift_scatter}
    \end{subfigure}
    \hfill
    \begin{subfigure}[t]{0.49\textwidth}
        \centering
        \includegraphics[width=\linewidth]{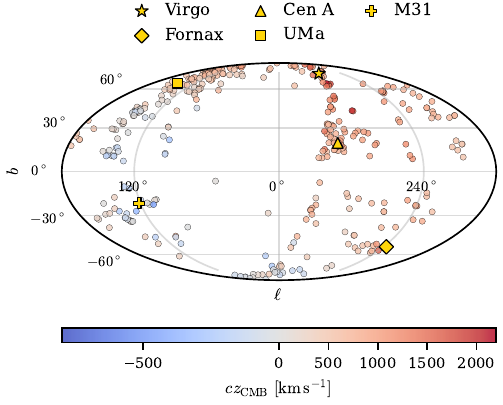}
        \caption{Sky distribution of the $401$ retained hosts in Galactic coordinates, coloured by $c\zCMB$.
        Yellow symbols mark Virgo, Fornax, Centaurus\,A, Ursa Major (UMa), and the M31 group.
        The sky distribution closely follows the local large-scale structure.}
        \label{fig:trgb_sky_distribution}
    \end{subfigure}
    \caption{Parent \ac{EDD} \ac{TRGB} sample of $445$ hosts with complete tabulated photometry and redshifts.}
    \label{fig:trgb_data_overview}
\end{figure*}

\Cref{fig:trgb_magnitude_redshift_scatter} shows the retained hosts in the joint $\mobs$--$c\zCMB$ plane, with marginal distributions for both observables.
\Cref{fig:trgb_sky_distribution} shows the sky distribution of the same hosts in Galactic coordinates, coloured by $c\zCMB$, with the directions of Virgo, Fornax, Ursa Major, Centaurus\,A, and M31 marked.
The hosts concentrate towards these nearby overdensities, reflecting the anisotropic large-scale structure of the local Universe.

We model the F814W tip absolute magnitude as a linear function of the tip colour $c$ about the pivot $c_\star$, following the \ac{EDD} colour standardisation of~\citet{Rizzi_2007},
\begin{equation}\label{eq:M_F814W_TRGB}
    M_{\rm F814W} = M_0 + \alpha_c\left(c - c_\star\right),
\end{equation}
where the zero-point $M_0$ and pivot $c_\star$ are inferred jointly with the other model parameters.
We fix the colour--magnitude slope to $\alpha_c=0.2$, the value adopted by~\citet{Anand_2022}.
Positive $\alpha_c$ means that redder tips are fainter in F814W.
Freeing $\alpha_c$ leaves it poorly constrained in our inference, so we keep it fixed at the \ac{EDD} value.
We do not use the \ac{EDD}-reported distance moduli within the inference.
Instead, we forward-model the tip magnitudes and redshifts to infer the distances as latent variables.
The catalogue does not report a redshift uncertainty.
We adopt a baseline value of $20\kmsec$, which is in any case subdominant to the expected peculiar-velocity scatter $\sigma_v$.
The retained rows define the per-host observed quantities $(\mobs,\,e_m,\,\cobs,\,e_c,\,c\zCMB,\,e_{cz})$ entering the likelihood.

We calibrate the \ac{TRGB} with two geometric anchors: the \ac{LMC} and NGC\,4258.
The \ac{LMC} has distance modulus $\mu_{\rm LMC}=18.477\pm0.026$ mag from detached eclipsing-binary distances~\citep{Pietrzynski_2019} and F814W tip magnitude $m_{\rm LMC}=14.456\pm0.018$ mag~\citep{Hoyt_2023}.
NGC\,4258 has distance modulus $\mu_{\rm N4258}=29.397\pm0.032$ mag from water-maser orbits~\citep{Reid_2019} and F814W tip magnitude $m_{\rm N4258}=25.347\pm0.044$ mag~\citep{Jang_2021}.
We identify the published anchor tip magnitudes with our $\mobs$ of~\cref{eq:mobs}.

\section{Inference framework}\label{sec:methods}

We construct a generative model in which the \ac{TRGB} calibration, geometric anchor distances, host redshifts, and catalogue selection are all forward-modelled together.
This requires a set of global parameters describing the population and a set of per-host latent variables.
We summarise the model in a directed acyclic graph in~\cref{fig:trgb_model_dag}, and tabulate all model parameters and their priors in~\cref{tab:trgb_priors} of Appendix~\ref{sec:model_priors}.
We describe each component below.

\begin{figure*}
    \centering
    \includegraphics[width=\textwidth]{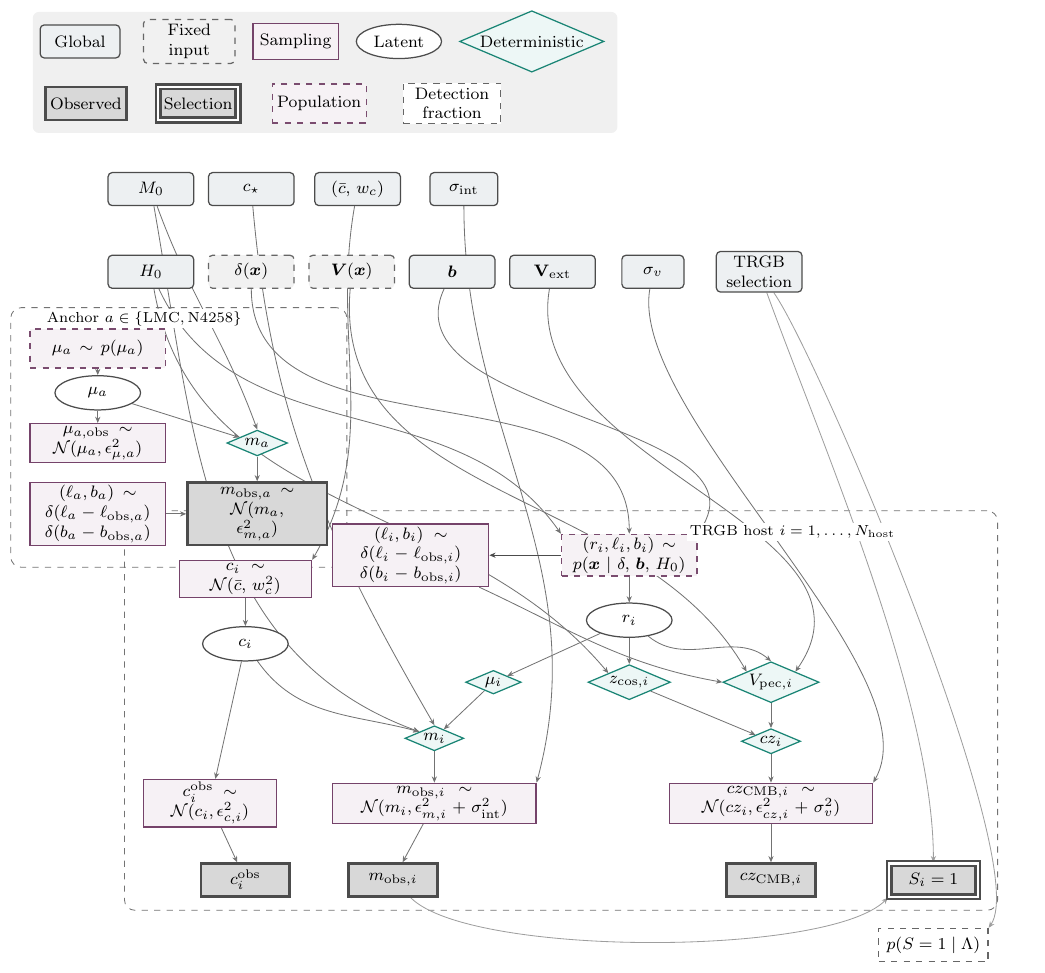}
    \caption{Directed acyclic graph of the hierarchical \ac{TRGB} model.
    The global parameters and the \Manticore\ fields generate the anchor and host observables through the latent anchor distance moduli $\mu_a$, the latent host distances $r_i$, and the latent tip colours $c_i$.
    The host angular coordinates $(\ell_i,\,b_i)$ are held at their observed values by delta-function nodes, so the \ac{LOS} distance is the only latent positional degree of freedom.
    The two plates repeat over the anchors $a\in\{{\rm LMC},\,{\rm NGC}\,4258\}$ and the \ac{TRGB} hosts.
    See~\cref{ssec:methods_global} for the generative model and~\cref{ssec:sel_model} for the selection treatment.
    }
    \label{fig:trgb_model_dag}
\end{figure*}

\subsection{Density and peculiar-velocity fields}\label{ssec:fields}

The inference requires external models of the density and velocity fields of the local Universe, for two reasons.
First, most hosts lie at $c\zCMB\lesssim 1000\kmsec$, where peculiar velocities of a few hundred $\kmsec$ are a substantial fraction of the recession velocity.
Second, the hosts trace the local large-scale structure through the galaxy bias rather than filling space uniformly, so the prior probability that a host lies at a given position depends on the matter density field (``inhomogeneous Malmquist bias'').
We therefore condition the inference on \Manticore~\citep{McAlpine_2025}, a reconstruction of the local density field $\delta(\bm{x})$ and peculiar-velocity field $\bm{V}(\bm{x})$.

\Manticore\ takes as input the \TWOMPP\ galaxy compilation of~\citet{Lavaux_2011}.
The compilation comprises $69{,}160$ galaxies from 2MASS~\citep{Skrutskie_2006} with redshifts from 2MRS~\citep{Huchra_2012}, 6dFGRS~\citep{Jones_2009}, and SDSS DR7~\citep{Abazajian_2009}.
It is $K_s$-limited at $11.5$ in the 2MRS region and $12.5$ elsewhere, with the usual photometric corrections applied.
\Manticore\ is a field-level inference of the local Universe with the \ac{BORG} algorithm~\citep{Jasche_2013,Lavaux_2016,Jasche_2019}.
Linear initial conditions at $z=1000$ are sampled by Hamiltonian Monte Carlo on a $256^3$ grid in a $681\Mpch$ box, giving an inference resolution of $2.66\Mpch$.
The galaxy-bias model is specified per magnitude and redshift bin, with a generalised-Poisson likelihood.
The initial conditions are evolved to $z=0$, with \texttt{COLA}, a comoving Lagrangian acceleration solver, simulating the gravitational evolution from $z=69$ onwards~\citep{COLA,Stopyra_2023}.
The inference adopts the Dark Energy Survey Year~3 ``$3\times2$pt~$+$~All Ext.'' \acs{LCDM} cosmology~\citep{DES_Y3}, with $h=0.681$ and $\Om=0.306$.
Because the field coordinates in \Mpch\ are set by the observed redshifts, we read the fields at $h\,r_i$ using the sampled $H_0$ rather than the fixed $h=0.681$.

In this work we use the $80$ posterior realisations evolved to $z=0$ with \texttt{COLA} at the native \ac{BORG} settings.
Their density and velocity fields are deposited on a $256^3$ grid with the piecewise-cubic-spline mass-assignment scheme.
In the baseline configuration we smooth the density field with a Gaussian kernel of width $4\Mpch$ and leave the peculiar-velocity field unsmoothed.
The constrained volume extends to $r\lesssim155\Mpch$, set by the \TWOMPP\ completeness.
\citet{McAlpine_2025} validate the reconstruction with posterior predictive tests of the matter power spectrum, the bispectrum, the halo mass function, and the Gaussianity of the inferred initial conditions, and with the recovery of the local cluster catalogue.
The velocity field is validated separately with the ``Velocity Field Olympics'' methodology~\citep{VF_olympics}.
We condition the inference on one realisation at a time and combine the $80$ resulting posteriors afterwards, under the two weightings defined in~\cref{ssec:field_marg}.

\subsection{Generative model}\label{ssec:methods_global}

The generative model has two per-host latent quantities, the \ac{LOS} distance $r_i$ and the true tip colour $c_i$, together with a global parameter vector $\bm{\Lambda}$ describing the population and shared model components.
Conditioned on $\bm{\Lambda}$ and the reconstructed overdensity field $\delta(\bm{x})$, the host-position prior is
\begin{equation}\label{eq:position_prior}
    \pi(\bm{r}_i \mid \bm{\Lambda},\,\delta(\bm{x}))
    =
    \frac{n(\bm{r}_i,\,\delta(\bm{x}),\,\bm{b})}
         {\int_{\mathcal{V}} {\rm d}^3\bm{r}\, n(\bm{r},\,\delta(\bm{x}),\,\bm{b})},
\end{equation}
where $\bm{b}$ denotes the galaxy-bias parameters, which control the mapping from the matter overdensity field to the intrinsic source number density.
This density weighting is the inhomogeneous Malmquist-bias term.
In~\cref{ssec:sel_model} we show that the normalisation of~\cref{eq:position_prior} need not be evaluated explicitly.
We treat the observed angular position as exact, equivalent to imposing the delta-function angular likelihood $\mathcal{L}(\hat{\bm{r}}_{i,{\rm obs}}\mid\bm{r}_i)=\delta_{\rm D}^{(2)}(\hat{\bm{r}}_i-\hat{\bm{r}}_{i,{\rm obs}})$.
Changing variables to spherical coordinates, ${\rm d}^3\bm{r}_i=r_i^2\,{\rm d}r_i\,{\rm d}\Omega$, and conditioning on $\hat{\bm{r}}_{i,{\rm obs}}$ gives
\begin{equation}\label{eq:distance_prior}
    \tilde{\pi}(r_i \mid \bm{\Lambda},\,\delta(\bm{x}),\,\hat{\bm{r}}_{i,{\rm obs}})
    =
    \frac{r_i^2\, n\!\left(\bm{r}_i,\,\delta(\bm{x}),\,\bm{b}\right)}
         {\int_{\mathcal{V}} {\rm d}^3\bm{r}\, n(\bm{r},\,\delta(\bm{x}),\,\bm{b})},
\end{equation}
where $\bm{r}_i=r_i\hat{\bm{r}}_{i,{\rm obs}}$ and the factor $r_i^2$ is the Jacobian of the coordinate transformation.
This ``volume prior'' weights each radial shell by the volume it subtends at fixed solid angle and is sometimes called the homogeneous Malmquist bias.%
\footnote{The same name is sometimes given to the apparent brightening of the mean absolute magnitude of a magnitude-selected sample, an effect that belongs to the selection function rather than to the distance prior.
We model it separately in~\cref{ssec:sel_model}.}
We therefore interpret~\cref{eq:distance_prior} as the \ac{LOS} representation of the same three-dimensional prior, not as an independently specified radial population model.
The tilde indicates that $\tilde{\pi}$ is unnormalised.
Its integral over $r_i$ gives the marginal probability of observing a source along that~\ac{LOS}, since the denominator in~\cref{eq:distance_prior} is the full three-dimensional volume integral.
For \Manticore\ we adopt the double-power-law density model
\begin{equation}\label{eq:bias_manticore}
\begin{split}
    n(\bm{x},\,\delta(\bm{x}),\,\bm{b})
    &\propto
    \left(\frac{1+\delta(\bm{x})}{\rho_t}\right)^{\alpha_{\rm low}}
    \\
    &\times
    \left[1 + \left(\frac{1+\delta(\bm{x})}{\rho_t}\right)^{1/\Delta_{\ln\rho}}\right]^{(\alpha_{\rm high}-\alpha_{\rm low})\Delta_{\ln\rho}},
\end{split}
\end{equation}
with $\bm{b}=\{\alpha_{\rm low},\,\alpha_{\rm high},\,\ln\rho_t,\,\Delta_{\ln\rho}\}$ controlling the low- and high-density slopes about the transition $1+\delta(\bm{x})=\rho_t$, and $\Delta_{\ln\rho}$ setting the transition width in $\ln\rho$.
We infer $\bm{b}$ jointly with the other global parameters.

We model the latent colour variable $c_i$ of the $i$\textsuperscript{th} host with a Gaussian hyperprior of mean $\bar c$ and width $w_c$, both inferred jointly with the other global parameters,
\begin{equation}\label{eq:colour_parent}
    \pi(c_i\mid\bar c,\,w_c) =
    \mathcal{N}\!\left(c_i;\,\bar c,\, w_c^2\right).
\end{equation}
The observed colour then enters through the Gaussian likelihood
\begin{equation}\label{eq:colour_obs_lik}
    \mathcal{L}(c_{{\rm obs},i}\mid c_i) =
    \mathcal{N}\!\left(c_{{\rm obs},i};\,c_i,\, e_{c,i}^2\right).
\end{equation}
The latent colour $c_i$ enters the tip absolute magnitude through~\cref{eq:M_F814W_TRGB}.
Combining this calibration with the distance modulus $\mu$ implied by the sampled distance gives the predicted apparent tip magnitude
\begin{equation}\label{eq:m_pred}
    m_i = M_0 + \alpha_c (c_i - c_\star) + \mu(r_i).
\end{equation}
The measured tip magnitude then enters as
\begin{equation}\label{eq:m_obs_lik}
    \mathcal{L}(m_{{\rm obs},i}\mid c_i,\,r_i,\,\bm{\Lambda}) =
    \mathcal{N}\!\left(m_{{\rm obs},i};\,m_i,\, e_{m,i}^2 + \sigma_{\rm int}^2\right),
\end{equation}
where the variance combines the F814W measurement uncertainty and the intrinsic \ac{TRGB} scatter $\sigma_{\rm int}$ in quadrature.
We infer $\sigma_{\rm int}$ jointly with the other global parameters.

The host \ac{LOS} peculiar velocity is predicted as
\begin{equation}\label{eq:vlos}
    V_{{\rm pec},i} =
    \hat{\bm{r}}_i \cdot
    \left[\bm{V}(\bm{r}_i) + \Vext\right],
\end{equation}
where $\bm{V}$ is the reconstructed peculiar-velocity field, adopted at face value, and $\Vext$ is an external bulk-flow vector modelling first-order coherent motion sourced outside the reconstructed volume and residual coherent errors within it.
The velocity-monopole variant examined in~\cref{ssec:results_flow} adds a spatially constant radial term $V_{\rm mono}$ in the \ac{CMB} frame to the predicted peculiar velocity, $V_{{\rm pec},i}\to V_{{\rm pec},i}+V_{\rm mono}$.
This term is fixed to zero in the baseline and freed only in that variant.
The predicted source redshift is then
\begin{equation}\label{eq:cz_pred}
    1 + z_i =
    \left[1 + z_{\rm cosmo}(r_i,\,H_0)\right]
    \left[1 + V_{{\rm pec},i}/c\right],
\end{equation}
where $H_0$ appears through the mapping from the sampled distance $r_i$ to the cosmological redshift $z_{\rm cosmo}$.
The observed redshift enters through the Gaussian likelihood
\begin{equation}\label{eq:cz_lik}
\begin{aligned}
    &\mathcal{L}(cz_{{\rm CMB},i}\mid r_i,\,\bm{V}(\bm{x}),\,\bm{\Lambda})
    \\
    &\qquad =
    \mathcal{N}\!\left(cz_{{\rm CMB},i};\,c z_i,\, e_{cz,i}^2+\sigma_v^2\right),
\end{aligned}
\end{equation}
where $e_{cz,i}$ is the redshift measurement uncertainty and $\sigma_v$ the residual redshift scatter, which we infer as a free parameter.
As a more outlier-robust alternative, we also consider the Student-$t$ redshift likelihood
\begin{equation}\label{eq:cz_student_lik}
\begin{aligned}
    \mathcal{L}(cz_{{\rm CMB},i}\mid r_i,\,\bm{V}(\bm{x}),\,\bm{\Lambda})
    &=
    \frac{\Gamma\!\left[(\nu+1)/2\right]}
    {\Gamma\!\left(\nu/2\right)\sqrt{\nu\pi}\,\sqrt{e_{cz,i}^2+\sigma_v^2}}
    \\
    &\times
    \left[1 + \frac{(cz_{{\rm CMB},i}-c z_i)^2}
    {\nu(e_{cz,i}^2+\sigma_v^2)}\right]^{-(\nu+1)/2},
\end{aligned}
\end{equation}
where $\nu$ is the number of degrees of freedom, inferred jointly with the other parameters, and the Gaussian likelihood is recovered as $\nu\to\infty$.

The two geometric anchors enter through the same absolute-magnitude calibration as the \ac{EDD} \ac{TRGB} hosts.
They have latent three-dimensional positions, with their observed angular coordinates fixed exactly, but do not enter the redshift likelihood.
For each anchor $a\in\{\mathrm{LMC},\,\mathrm{NGC}\,4258\}$, we use a uniform-in-volume distance prior, $\pi(\bm{r}_a)\propto 1$,
rather than evaluating the inhomogeneous-Malmquist factor of~\cref{eq:position_prior}.
This is inconsequential because the geometric distance likelihood is narrow compared with the scale over which the density field varies, so the weighting is effectively constant across the distance range allowed by the geometric calibration.
The baseline density smoothing is $4\Mpch$, whereas the distance error of NGC\,4258 is only ${\sim}0.08\Mpch$ and the \ac{LMC} lies at ${\sim}0.03\Mpch$.
With $\mu_a\equiv\mu(r_a)$, the geometric distance measurement enters as
\begin{equation}\label{eq:anchor_mu_lik}
    \mathcal{L}(\mu_{a,{\rm obs}}\mid\mu_a) =
    \mathcal{N}\!\left(\mu_{a,{\rm obs}};\,\mu_a,\, e_{\mu,a}^2\right),
\end{equation}
where $\mu_{a,{\rm obs}}$ and $e_{\mu,a}$ are the reported geometric distance modulus and its uncertainty for anchor $a$.
We assume the F814W tip magnitude of each anchor is extinction-corrected and standardised to the pivot colour $c_\star$, so the anchor tip likelihood is
\begin{equation}\label{eq:anchor_m_lik}
\begin{aligned}
    &\mathcal{L}(m_{a,{\rm obs}}\mid\mu_a,\,M_0,\,\sigma_{\rm int})
    \\
    &\qquad =
    \mathcal{N}\!\left(m_{a,{\rm obs}};\,M_0+\mu_a,\, e_{m,a}^2+\sigma_{\rm int}^2\right).
\end{aligned}
\end{equation}
Assuming the two anchors are conditionally independent, we take the product of the two geometric-anchor likelihoods,
\begin{equation}\label{eq:anchor_likelihood}
\begin{aligned}
    \mathcal{L}_{\rm anc}(\bm{d}_a \mid \bm{\Lambda})
    &\equiv \prod_{a} \mathcal{L}(\mu_{a,{\rm obs}}\mid\mu_a) \\
    &\quad\times \mathcal{L}(m_{a,{\rm obs}}\mid\mu_a,\,M_0,\,\sigma_{\rm int}),
\end{aligned}
\end{equation}
where $\bm{d}_a$ denotes the anchor data.
For ease of notation, we include the sampled anchor distances $\mu_{\rm LMC}$ and $\mu_{\rm N4258}$ in the global parameter vector $\bm{\Lambda}$.

The latent host colour and distance can be sampled explicitly, but we marginalise over both to reduce the dimensionality of the posterior.
At fixed distance $r_i$, we analytically marginalise over the latent true colour $c_i$ to obtain the colour-marginal likelihood,
\begin{equation}\label{eq:colour_marginal}
    \begin{aligned}
    \bar{\mathcal{L}}(m_{{\rm obs},i}\mid r_i,\,\bm{\Lambda})
    &=
    \int {\rm d}c_i\,
    \pi(c_i\mid\bar c,\,w_c)\,
    \mathcal{L}(c_{{\rm obs},i}\mid c_i)
    \\
    &\quad\times
    \mathcal{L}(m_{{\rm obs},i}\mid c_i,\,r_i,\,\bm{\Lambda}).
    \end{aligned}
\end{equation}
Because the parent colour distribution, colour likelihood, and magnitude likelihood are Gaussian, and the predicted magnitude is linear in $c_i$, this integral evaluates to
\begin{equation}\label{eq:mc_lik}
    \begin{aligned}
    \bar{\mathcal{L}}(m_{{\rm obs},i}\mid r_i,\,\bm{\Lambda})
    &=
    \mathcal{N}\!\left(c_{{\rm obs},i};\,\bar c,\,w_c^2+e_{c,i}^2\right)
    \\
    &\quad\times
    \mathcal{N}\!\left(m_{{\rm obs},i};\,\bar m_i,\,\bar\sigma_{m,i}^2\right),
    \end{aligned}
\end{equation}
where
\begin{align}
    \tilde c_i
    &\equiv
    \frac{e_{c,i}^2\bar c + w_c^2 c_{{\rm obs},i}}
    {w_c^2 + e_{c,i}^2}, \label{eq:colour_post_mean}\\
    \tilde\sigma_{c,i}^2
    &\equiv
    \frac{w_c^2 e_{c,i}^2}
    {w_c^2 + e_{c,i}^2}, \label{eq:colour_post_var}\\
    \bar m_i
    &\equiv M_0 + \alpha_c(\tilde c_i - c_\star) + \mu(r_i), \\
    \bar\sigma_{m,i}^2
    &\equiv e_{m,i}^2 + \sigma_{\rm int}^2 + \alpha_c^2\,\tilde\sigma_{c,i}^2.
\end{align}
The first Gaussian in~\cref{eq:mc_lik} is the marginal likelihood of the observed colour under the parent population, while the second is the magnitude likelihood after propagating the latent-colour uncertainty.
Putting this all together, the distance- and colour-marginalised magnitude--redshift likelihood of a \ac{TRGB} host is
\begin{equation}\label{eq:host_mz_likelihood}
\begin{aligned}
    \mathcal{L}^{(k)}(\bm{d}_i \mid \bm{\Lambda})
    &\equiv
    \int {\rm d}r_i\,
    \tilde{\pi}(r_i \mid \bm{\Lambda},\,\delta^{(k)}(\bm{x}),\,\hat{\bm{r}}_{i,{\rm obs}})
\\
    &\quad\times
    \bar{\mathcal{L}}(m_{{\rm obs},i}\mid r_i,\,\bm{\Lambda})
    \\
    &\quad\times
    \mathcal{L}(cz_{{\rm CMB},i}\mid r_i,\,\bm{V}^{(k)}(\bm{x}),\,\bm{\Lambda}),
\end{aligned}
\end{equation}
where $\delta^{(k)}(\bm{x})$ and $\bm{V}^{(k)}(\bm{x})$ are the density and velocity fields of the $k$\textsuperscript{th} reconstructed field realisation.
We evaluate~\cref{eq:host_mz_likelihood} by Simpson quadrature\footnote{The radial grid spans $0.15$--$35\Mpc$ with $2007$ nodes, uniformly spaced in distance modulus by $0.00625$ mag below ${\sim}17\Mpc$ and in distance by $0.05\Mpc$ above.}, leaving only the global parameters to be sampled.
Given $\bm{\Lambda}$ and a fixed density- and velocity-field realisation, we assume the \ac{TRGB} hosts and geometric anchors are conditionally independent, so the joint likelihood is the product of the anchor likelihood in~\cref{eq:anchor_likelihood} and the per-host factors in~\cref{eq:host_mz_likelihood}.

\subsection{Selection modelling}\label{ssec:sel_model}

The model of~\cref{ssec:methods_global} describes the parent population and does not yet account for the selection effects that make the observed catalogue a non-random subset of it.
Let $\mathcal{A}_{\rm obs}$ denote the set of observed hosts.
Let $S_i=1$ denote inclusion of the $i$\textsuperscript{th} host in the observed catalogue, and define $\mathcal{S}_{\rm obs}\equiv\{S_i=1:\,i\in\mathcal{A}_{\rm obs}\}$.
Following our treatment in~\citet{Stiskalek_Cepheids,Stiskalek_2026,Stiskalek_2026_erratum} (itself following~\citealt{Kelly_2007,Kelly_2008}), the posterior conditioned on the detected catalogue and on the $k$\textsuperscript{th} field realisation is
\begin{equation}\label{eq:detected_posterior}
\begin{aligned}
    &\mathcal{P}^{(k)}(\bm{\Lambda} \mid \bm{d}_{\rm obs},\,\mathcal{S}_{\rm obs})
    \propto
    \pi(\bm{\Lambda})\,
    \mathcal{L}_{\rm anc}(\bm{d}_a \mid \bm{\Lambda})\,
    \\
    &\qquad\times
    \left[\prod_{i\in\mathcal{A}_{\rm obs}}
    p(S_i=1\mid m_{{\rm obs},i},\,\bm{\Lambda})\right]
    \\
    &\qquad\times
    \left[S^{(k)}(\bm{\Lambda})\right]^{-n}
    \prod_{i\in\mathcal{A}_{\rm obs}}
    \mathcal{L}^{(k)}(\bm{d}_i \mid \bm{\Lambda}),
\end{aligned}
\end{equation}
where $\pi(\bm{\Lambda})$ is the prior on the global parameters, $\bm{d}_{\rm obs}=\{\bm{d}_a,\,\{\bm{d}_i\}_{i\in\mathcal{A}_{\rm obs}}\}$ collects the anchor data $\bm{d}_a$ and the per-host \ac{TRGB} data $\bm{d}_i=(m_{{\rm obs},i},\,c_{{\rm obs},i},\,cz_{{\rm CMB},i})$, $n=|\mathcal{A}_{\rm obs}|$ is the number of observed hosts, and
$S^{(k)}(\bm{\Lambda})\equiv p(S=1\mid\bm{\Lambda},\,\delta^{(k)}(\bm{x}))$ is the population-detected fraction of the $k$\textsuperscript{th} \Manticore\ realisation, defined in~\cref{eq:S_volume}.

The \ac{EDD} sample has no single well-defined observing strategy, so the selection model cannot be constructed from a survey definition and must instead be posited parametrically.
We assume that selection acts on the tip apparent magnitude, i.e.\ that the per-host inclusion factor in~\cref{eq:detected_posterior} depends only on $m_{{\rm obs},i}$ through $p(S_i=1\mid m_{{\rm obs},i},\,\bm{\Lambda})$.
Measuring a tip requires imaging that resolves individual red giants and reaches fainter than the tip, so we take the apparent magnitude of the tip to be the observable that determines whether a host can enter the catalogue at all.

The other observable on which the selection could act is the redshift.
The parent list does not support such a cut: the Updated Nearby Galaxy Catalog of~\citet{Karachentsev_2013} selects a galaxy if $V_{\rm LG}<600\kmsec$ \emph{or} $D<11\Mpc$.
Because either condition alone suffices, the parent list truncates neither $c\zCMB$ nor $\mobs$, so no selection term of either form follows from it.
The magnitude window we adopt below stands as a posited proxy for the depth of the archival imaging.
A secondary dependence is on host properties: the dominant \ac{HST} programme targets unobscured, bright galaxies, and tip detectability depends further on star-formation history and crowding~\citep{Anand_2021}.
None of these reduces to a simple $\mobs$ selection.
We further restrict the observable sky with a Galactic-plane mask and allow an angular sky-exposure extension (\cref{sec:angular_selection}).
The posterior predictive checks of~\cref{sec:results_ppc} test a posteriori whether these assumptions, together with the generative model, reproduce the observed data.

For an observed apparent tip magnitude, we model the detection probability as a finite window between a fixed bright threshold $m_{\rm min}=22.1$ mag and an inferred faint threshold $\mlim$, each smoothed with the standard-normal cumulative distribution function $\Phi$ and a common width $\sigma_{\rm sel}$, so that
\begin{equation}\label{eq:p_sel}
    \begin{split}
    p(&S_i=1 \mid m_{{\rm obs},i},\,\bm{\Lambda}) =
    \\
    &\Phi\!\left(\frac{\mlim - m_{{\rm obs},i}}{\sigma_{\rm sel}}\right) -
    \Phi\!\left(\frac{m_{\rm min} - m_{{\rm obs},i}}{\sigma_{\rm sel}}\right).
    \end{split}
\end{equation}
The bright edge at $m_{\rm min}$ models the fixed $\mobs>22.1$ mag truncation, while the faint edge $\mlim$ captures the \ac{HST} magnitude limit.
A hard cut is precluded by the marginal F814W tip-magnitude distribution in~\cref{fig:trgb_magnitude_redshift_scatter}, which falls sharply near $26$ mag but has a tail of hosts at fainter magnitudes.
The faint threshold $\mlim$ and width $\sigma_{\rm sel}$ are global parameters in $\bm{\Lambda}$, while $m_{\rm min}$ is held fixed at the truncation value.
We evaluate the population-detected fraction per realisation, marginalising over the latent host position with the three-dimensional prior of~\cref{eq:position_prior},
\begin{equation}\label{eq:S_volume}
    S^{(k)}(\bm{\Lambda}) =
    \int_{\mathcal{V}} {\rm d}^3\bm{r}\,\pi(\bm{r}\mid\bm{\Lambda},\,\delta^{(k)}(\bm{x}))\,
    p(S=1\mid\bm{r},\,\bm{\Lambda}).
\end{equation}
The position-conditioned detection probability is obtained by marginalising~\cref{eq:p_sel} over the Gaussian likelihood of $\mobs$ at $\bm{r}$,
\begin{equation}\label{eq:p_sel_marg}
\begin{aligned}
    p(S=1\mid\bm{r},\,\bm{\Lambda})
    &=
    \int {\rm d}m_{\rm obs}\,
    \mathcal{N}\!\left(m_{\rm obs};\,\bar m(\bm{r};\bm{\Lambda}),\,\sigma_m^2\right)
    \\
    &\times
    \left[
    \Phi\!\left(\frac{\mlim - m_{\rm obs}}{\sigma_{\rm sel}}\right)
    -
    \Phi\!\left(\frac{m_{\rm min} - m_{\rm obs}}{\sigma_{\rm sel}}\right)
    \right]
    \\
    &=
    \Phi\!\left(\frac{\mlim - \bar m}{\sqrt{\sigma_{\rm sel}^2 + \sigma_m^2}}\right)
    -
    \Phi\!\left(\frac{m_{\rm min} - \bar m}{\sqrt{\sigma_{\rm sel}^2 + \sigma_m^2}}\right).
\end{aligned}
\end{equation}
Here $\bar m(\bm{r};\bm{\Lambda}) \equiv M_0+\alpha_c(\bar c-c_\star)+\mu(\bm{r};H_0)$ is the mean of the Gaussian likelihood and $\sigma_m^2 \equiv \sigma_{\rm int}^2+\langle e_m\rangle^2+\alpha_c^2 w_c^2$ its variance, propagating the intrinsic \ac{TRGB} scatter, a representative F814W measurement uncertainty, and the parent colour width through the colour--magnitude relation of~\cref{eq:m_pred}.
The normalisation $Z^{(k)}\equiv\int_{\mathcal{V}} {\rm d}^3\bm{r}\,n(\bm{r},\,\delta^{(k)}(\bm{x}),\,\bm{b})$ of~\cref{eq:position_prior} cancels realisation-by-realisation between the $n$ powers in $[S^{(k)}(\bm{\Lambda})]^{-n}$ and the matching denominator of each of the $n$ per-host likelihoods via~\cref{eq:distance_prior}.
We therefore never evaluate $Z^{(k)}$ and need only the volume integral of~\cref{eq:S_volume} with the unnormalised position prior.
The overall normalisation of $p(S=1\mid m)$ cancels in the same way between the per-host inclusion factors and $[S^{(k)}(\bm{\Lambda})]^{-n}$, so only the shape of the selection window is constrained and the detection probability need not reach unity anywhere.
Because the observed angular coordinate enters as a delta-function angular likelihood, the host term collapses onto the observed \ac{LOS}, whereas the selection term remains an integral over the full survey volume.
For no-reconstruction runs the same convention gives the full-sky radial selection normalisation $4\pi\int {\rm d}r\,r^2p(S=1\mid r,\,\bm{\Lambda})$.

Since the \Manticore\ density field is defined on a regular grid, we evaluate the volume integral as a sum over $\mathcal{V}$, a $75\Mpc$ sphere around the observer.
This choice is sufficient because the $p(S=1\mid\bm{r},\,\bm{\Lambda})$ term assigns negligible weight to any volume outside it.
The selection integrand has support only to ${\sim}\,10\Mpc$, comparable to the $3.9\Mpc$ voxel size of the \Manticore\ fields, so a direct voxel sum would render the integral noisy.
We therefore super-resolve every voxel within ${\sim}20\Mpc$ by splitting it into $512$ subcells of spacing ${\sim}\,0.5\Mpc$, evaluating the fields at the subcell centres by trilinear interpolation and dividing the parent volume weight among them.
Because we model the selection as a function of magnitude, it is independent of the velocity field, $\Vext$, and $\sigma_v$.

In principle we should marginalise over the per-host distribution of $e_m$ alongside the selection variance $\sigma_{\rm sel}$.
Instead, we approximate this by replacing the per-host $e_{m,i}$ values with the median magnitude uncertainty $\langle e_m\rangle$ in $\sigma_m$.
Since $\sigma_m$ and $\sigma_{\rm sel}$ enter in quadrature, and the latter is much larger, this approximation has a negligible effect on the results.

\subsubsection{Angular selection}\label{sec:angular_selection}

Two angular terms enter the selection model in addition to the magnitude window.
First, we restrict to Galactic latitudes $|b|\geq b_{\rm min}$ with $b_{\rm min}=10^\circ$, an approximate Zone-of-Avoidance mask reflecting the absence of \ac{TRGB} data where extinction prevents their measurement~\citep{KraanKorteweg_2000, KraanKorteweg_2005}.
The mask multiplies the integrand of~\cref{eq:S_volume} by an indicator function, so the population-detected fraction is evaluated over the unmasked volume only.

Second, in the variants labelled ``sky exposure'' we allow the detection probability to depend on direction as well, since the \ac{HST} pointing history does not sample the sky uniformly.
In these variants we partition the sky into $N_{\rm pix}$ equal-area \texttt{HEALPix} pixels~\citep{Gorski_2005}, indexed by $q$, and assign each a direction-dependent exposure weight $\theta_q$.
We test an $N_{\rm side}=1$ partition ($N_{\rm pix}=12$) and a finer $N_{\rm side}=2$ partition ($N_{\rm pix}=48$).
The weight enters the selection in two places: it multiplies the magnitude term of~\cref{eq:p_sel} for a host lying in the $q$\textsuperscript{th} pixel, and it weights the corresponding contribution to the volume integral.
The population-detected fraction of~\cref{eq:S_volume} then becomes the exposure-weighted sum $S^{(k)}(\bm{\Lambda})\propto\sum_q \theta_q\, S_q^{(k)}(\bm{\Lambda})$ over the per-pixel selection integrals.

We set a symmetric Dirichlet prior on the exposure weights, $\bm{\theta}\hookleftarrow\mathrm{Dirichlet}(\alpha\bm{1})$, with the concentration fixed at $\alpha=4$ across both resolutions.
The total concentration $\kappa\equiv\alpha N_{\rm pix}$ is then $48$ at $N_{\rm side}=1$ and $192$ at $N_{\rm side}=2$.
Uniform exposure would be recovered as $\alpha\to\infty$, with $\alpha\to0$ the opposite, maximally uneven limit.
The normalisation cancellation of~\cref{ssec:sel_model} means that only the exposure contrasts between pixels are constrained.
This choice is weakly informative: each weight then has prior mean $1/N_{\rm pix}$ and fractional scatter $\sqrt{(N_{\rm pix}-1)/(\kappa+1)}\approx0.5$ at both resolutions.
The prior therefore admits factor-of-a-few pixel-to-pixel contrasts while disfavouring configurations in which nearly all of the exposure concentrates in one pixel.

\subsection{Field marginalisation}\label{ssec:field_marg}

The \Manticore\ posterior over the density and velocity fields is represented by an ensemble of $N_{\rm field}=80$ samples $\{\delta^{(k)}(\bm{x}),\,\bm{V}^{(k)}(\bm{x})\}$, over which the field must be marginalised.
We write $\bm{F}\equiv\{\delta(\bm{x}),\,\bm{V}(\bm{x})\}$ for a single realisation and $\mathcal{P}(\bm{F})$ for the distribution the ensemble represents.
With $\mathcal{L}(\bm{d}_{\rm obs},\,\mathcal{S}_{\rm obs}\mid\bm{\Lambda},\,\bm{F})$ of~\cref{eq:detected_posterior}, the field-marginalised posterior is
\begin{equation}\label{eq:field_marginal}
\begin{aligned}
    \mathcal{P}_{\rm ev}(\bm{\Lambda})
    &=
    \frac{1}{\mathcal{Z}}\int\dd\bm{F}\,
    \mathcal{L}(\bm{d}_{\rm obs},\,\mathcal{S}_{\rm obs}\mid\bm{\Lambda},\,\bm{F})\,
    \pi(\bm{\Lambda})\,\mathcal{P}(\bm{F})
    \\
    &=
    \frac{1}{\mathcal{Z}}\int\dd\bm{F}\,
    \mathcal{P}(\bm{\Lambda}\mid\bm{d}_{\rm obs},\,\mathcal{S}_{\rm obs},\,\bm{F})\,
    \mathcal{Z}(\bm{F})\,\mathcal{P}(\bm{F}),
\end{aligned}
\end{equation}
where $\mathcal{Z}(\bm{F})\equiv\int\dd\bm{\Lambda}\,\mathcal{L}(\bm{d}_{\rm obs},\,\mathcal{S}_{\rm obs}\mid\bm{\Lambda},\,\bm{F})\,\pi(\bm{\Lambda})$ is the single-field evidence, $\mathcal{P}(\bm{\Lambda}\mid\bm{d}_{\rm obs},\,\mathcal{S}_{\rm obs},\,\bm{F})$ the single-field posterior of~\cref{eq:detected_posterior}, and $\mathcal{Z}=\int\dd\bm{F}\,\mathcal{Z}(\bm{F})\,\mathcal{P}(\bm{F})$ the field-marginalised evidence.
The second line follows by applying Bayes' theorem.

The $N_{\rm field}$ realisations are equally weighted draws from $\mathcal{P}(\bm{F})$, so the field integrals of~\cref{eq:field_marginal} reduce to unweighted sums over them.
We write $\mathcal{P}^{(k)}(\bm{\Lambda})\equiv\mathcal{P}(\bm{\Lambda}\mid\bm{d}_{\rm obs},\,\mathcal{S}_{\rm obs},\,\bm{F}^{(k)})$ and $\mathcal{Z}^{(k)}\equiv\mathcal{Z}(\bm{F}^{(k)})$ for the posterior and evidence of the $k$\textsuperscript{th} realisation.
The second line of~\cref{eq:field_marginal} is then estimated by $(1/N_{\rm field})\sum_k\mathcal{Z}^{(k)}\,\mathcal{P}^{(k)}(\bm{\Lambda})$ and its normalisation by $\mathcal{Z}\simeq(1/N_{\rm field})\sum_k\mathcal{Z}^{(k)}$.
The common factor of $1/N_{\rm field}$ cancels in the ratio, leaving the evidence-weighted mixture
\begin{equation}\label{eq:evidence_stack}
    \mathcal{P}_{\rm ev}(\bm{\Lambda})
    \simeq
    \frac{\sum_k \mathcal{Z}^{(k)}\,\mathcal{P}^{(k)}(\bm{\Lambda})}
         {\sum_k \mathcal{Z}^{(k)}}.
\end{equation}
The evidence $\mathcal{Z}^{(k)}$ quantifies how well a given density--velocity field explains the data.
We sample~\cref{eq:detected_posterior} separately for each realisation and combine the $80$ posteriors according to~\cref{eq:evidence_stack} afterwards.
We call this the evidence-weighted treatment.

We define the effective number of contributing realisations as
\begin{equation}
    N_{\rm eff}
    \equiv
    \frac{\left(\sum_k \mathcal{Z}^{(k)}\right)^2}{\sum_k\left(\mathcal{Z}^{(k)}\right)^2},
\end{equation}
which is Kish's effective sample size $1/\sum_k w_k^2$ of the evidence weights $w_k\equiv \mathcal{Z}^{(k)}/\sum_j \mathcal{Z}^{(j)}$~\citep{Kish_1965}.
It runs from $N_{\rm field}$ for uniform weights down to unity when a single realisation dominates.
We also report the equal-weight stack
\begin{equation}\label{eq:equal_stack}
    \mathcal{P}_{\rm eq}(\bm{\Lambda})
    =
    \frac{1}{N_{\rm field}}\sum_k\mathcal{P}^{(k)}(\bm{\Lambda}),
\end{equation}
which weights all $80$ realisations equally.
This is not a field marginalisation but what follows if every realisation is equally plausible.
We report it because its width retains the scatter between realisations, which the evidence-weighted mixture loses when a few realisations dominate the weights.

\Cref{eq:field_marginal} shows that averaging the likelihood over the ensemble within a single sampler chain targets the same mixture as~\cref{eq:evidence_stack}, i.e.\ Bayesian model averaging of the $80$ realisations.
The two treatments therefore differ only in how that mixture is sampled.
The mixture of~\cref{eq:evidence_stack} is multimodal in $\bm{\Lambda}$ whenever the posteriors of individual realisations peak at different $\bm{\Lambda}$, in which case the sampler may fail to mix between modes and remain in whichever one the warm-up adapts to.
We find this to be the case here, and therefore sample one field at a time and weight the fields either by their evidence or equally afterwards.
This also makes the weight of each realisation and the field-by-field scatter explicit.

\subsection{Inference configuration and model comparison}\label{ssec:inference_config}
For all terms that require a background distance--redshift relation, we use a fixed flat $\Om=0.3$ cosmology to convert $r_i$ into the host distance modulus $\mu_i$ and cosmological redshift $z_{{\rm cosmo},i}$.
Given that the \ac{TRGB} hosts are all at $z<0.01$, any reasonable variation in $\Om$ has a negligible effect on the inference.

We sample the posterior with the No-U-Turn Sampler as implemented in \texttt{NumPyro}\footnote{\url{https://num.pyro.ai/en/stable/}}~\citep{Hoffman_2011,Phan_2019}, using five chains with $2000$ warm-up steps and $5000$ posterior samples, and require $\hat R<1.01$ for all sampled parameters~\citep{Gelman_1992,Vehtari_2021}.
We compare model variants through the Bayesian evidence, estimated from the posterior samples with the normalising-flow-learnt harmonic-mean estimator of the \texttt{harmonic} package~\citep{McEwen_2021,Mancini_2023,Polanska_2023,Piras_2024}.
Its numerical error has a median of $0.44$ in $\logZ$ per realisation, which is subdominant to the typical evidence differences we interpret below.
We report evidence differences as $\Delta\logZ$, quoted in~\cref{tab:trgb_h0_variants} separately for each statistic relative to the highest-scoring variant in the table, so that the preferred variant has $\Delta\logZ=0$ and the remaining variants are negative.
Where two models are compared directly in the text, we instead quote the difference between them and name the model it favours.

We define the ensemble evidence of a model as the mean of its single-field evidences,
\begin{equation}\label{eq:z_ens}
    \mathcal{Z}_{\rm ens}
    \equiv
    \frac{1}{N_{\rm field}}\sum_k \mathcal{Z}^{(k)}.
\end{equation}
This is the Monte Carlo estimate of the field-marginalised evidence $\mathcal{Z}$ of~\cref{eq:field_marginal}, so it is the evidence of the model after marginalising over the \Manticore\ ensemble.
We compare model $A$ with model $B$ by either of two evidence differences,
\begin{align}
    \Delta\logZ_{\rm ens}
    &=
    \log\frac{\sum_k \mathcal{Z}_A^{(k)}}{\sum_k \mathcal{Z}_B^{(k)}},
    \label{eq:dlogz_ens}
    \\
    \Delta\logZ_{\rm field}
    &=
    \frac{1}{N_{\rm field}}\sum_k\log\frac{\mathcal{Z}_A^{(k)}}{\mathcal{Z}_B^{(k)}},
    \label{eq:dlogz_field}
\end{align}
where $\mathcal{Z}_A^{(k)}$ and $\mathcal{Z}_B^{(k)}$ are the evidences of the two models on the $k$\textsuperscript{th} realisation.
While $\Delta\logZ_{\rm ens}$ is the log Bayes factor between the two models after marginalising over the field realisations, $\Delta\logZ_{\rm field}$ is the mean of the per-realisation log evidence ratio.
The former is thus dominated by the highest-evidence realisations, whereas the latter weights all realisations equally.
\Cref{tab:trgb_h0_variants} reports both statistics for each model variant, quoting $\Delta\logZ_{\rm field}$ along with the standard deviation of the log evidence ratios across realisations,
\begin{equation}\label{eq:dlogz_field_std}
    \sigma_{\rm field}
    \equiv
    \left[\frac{1}{N_{\rm field}}\sum_k\left(\log\frac{\mathcal{Z}_A^{(k)}}{\mathcal{Z}_B^{(k)}}-\Delta\logZ_{\rm field}\right)^2\right]^{1/2}.
\end{equation}

\section{Results}\label{sec:results}

We first present the baseline inference~(\cref{ssec:results_baseline}) and test it against the data with posterior predictive checks~(\cref{sec:results_ppc}).
We then quantify how the inference depends on each modelling choice in turn: the redshift likelihood~(\cref{ssec:results_zlike}), the density-field smoothing scale~(\cref{ssec:results_smoothing}), the angular sky-exposure model~(\cref{ssec:results_skyexp}), and the flow modelling~(\cref{ssec:results_flow}).
Throughout we quote each result under the two treatments of the \Manticore\ realisations defined in~\cref{ssec:field_marg}, the evidence-weighted and the equal-weight.
We compare variants by the evidence as set out in~\cref{ssec:inference_config}.
The results are tabulated in~\cref{tab:trgb_h0_variants}, which reports the two treatments side by side for each variant.

\subsection{Baseline inference}\label{ssec:results_baseline}

\begin{figure}
    \centering
    \includegraphics[width=\columnwidth]{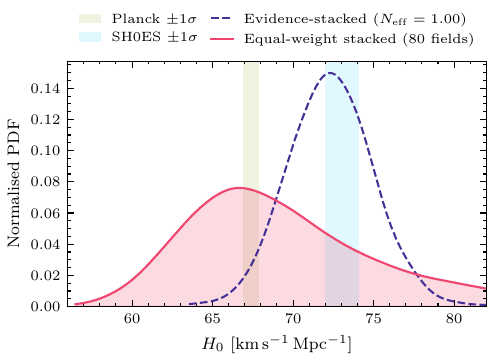}
    \caption{Marginal $H_0$ posteriors of the baseline \Manticore\ inference: the evidence-weighted posterior of~\cref{eq:evidence_stack} (dashed) versus the equal-weight stack of the same $80$ single-field posteriors (solid, filled), against the $\pm1\sigma$ bands of \ac{SHOES}~\citep{Riess_2022} and \textit{Planck} \acs{LCDM}~\citep{Planck_2020_cosmo}.
    The evidence-weighted posterior peaks near $72\kmsecMpc$ but has collapsed onto a single \Manticore\ realisation.
    }
    \label{fig:trgb_h0_comparison}
\end{figure}

\begin{figure}
    \centering
    \includegraphics[width=\columnwidth]{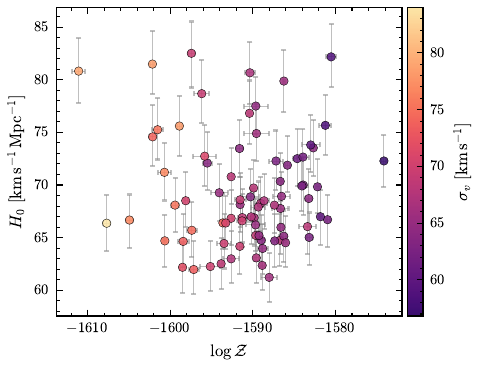}
    \caption{Per-field $H_0$ against the single-field evidence $\logZ$ for the baseline \Manticore\ model ($4\Mpch$ density smoothing, a Student-$t$ redshift likelihood, and the $48$-pixel angular sky-exposure selection term) with error bars denoting the $1\sigma$ posterior uncertainty of each single-field run.
    Points are coloured by the inferred Student-$t$ scale $\sigma_v$.
    A single realisation dominates the evidence and lies at $H_0\simeq72\kmsecMpc$, while the $80$ field medians scatter by ${\sim}5\kmsecMpc$, so the inference remains highly sensitive to the reconstruction realisation.}
    \label{fig:trgb_baseline_field_diagnostics}
\end{figure}

We select the baseline variant by its ensemble evidence $\mathcal{Z}_{\rm ens}$.
It is dominated by a single realisation, as we show below, so we check each ranking against the equally weighted $\Delta\logZ_{\rm field}$.
The two typically agree.
The baseline model is the \Manticore\ reconstruction with $4\Mpch$ density smoothing, a Student-$t$ redshift likelihood, and the $48$-pixel angular sky-exposure selection term.
The evidence-weighted constraint is $H_0=72.3\pm2.6\kmsecMpc$~(\cref{tab:trgb_h0_variants}).
Stacking the same per-field posteriors with equal weights instead gives $H_0=68.4\pm5.9\kmsecMpc$, and~\cref{fig:trgb_h0_comparison} shows the baseline $H_0$ posterior under both treatments.
The $80$ field medians scatter about the equal-weight median with a standard deviation of $5.2\kmsecMpc$~(\cref{fig:trgb_baseline_field_diagnostics}).
The $5.9\kmsecMpc$ equal-weight width is essentially the quadrature sum of this scatter and the ${\sim}2.6\kmsecMpc$ mean single-field uncertainty.
The field-to-field scatter therefore limits the equal-weight constraint, not the per-field statistical precision.
The evidence shows no significant preference between this baseline and the velocity-monopole variant.
That variant gives an $H_0$ consistent with the baseline under the evidence-weighted treatment and higher by $1\sigma$ under the equal-weight treatment~(\cref{ssec:results_flow}).

The evidence-weighted posterior of~\cref{eq:evidence_stack} has $N_{\rm eff}=1.00$ out of $80$.
The single-field evidences scatter by $6.7$ in $\logZ$, and the dominant realisation exceeds the runner-up by $\Delta\logZ=6.4$.
The four next-best realisations have field medians between $66.7$ and $82.2\kmsecMpc$.
The physical reason is that the \ac{TRGB} hosts mostly lie at $c\zCMB\lesssim1000\kmsec$ and each probes $\delta$ and $\bm{V}$ on the small scales that the \TWOMPP\ constraints determine least well.
Hence the $80$ \Manticore\ realisations, which agree on the large-scale flow, still differ slightly in how well they explain these redshifts.
Since the likelihood is a product over the $394$ hosts, these small per-host differences compound, making the realisations meaningfully different.
An ensemble of $80$ is therefore too small to marginalise the field out, since the dominant realisation would likely differ for a different set of draws.
The evidence-weighted stack is the marginal over the reconstruction posterior, and its limitation is that we approximate this marginal with only $80$ samples.
The equal-weight stack is not a marginalisation but is what follows if all $80$ realisations are equally plausible, so the two together bracket the sensitivity to the reconstruction.

The \ac{TRGB} calibration and selection parameters remain unchanged across the model variants.
The tip zero-point is $M_0=-4.03\pm0.08$ mag and the intrinsic scatter $\sigma_{\rm int}=0.10\pm0.01$ mag, recovered within these uncertainties by all model variants.
We note that we impose an informative prior on $\sigma_{\rm int}$.
It is obtained by inferring the \ac{TRGB} distances from the tip magnitudes alone, calibrated on the same \ac{LMC} and NGC\,4258 geometric anchors as in the rest of this work, but without the density and velocity fields or redshift information and hence without inferring $H_0$.
The prior breaks the partial degeneracy between $\sigma_{\rm int}$ and the peculiar-velocity scatter $\sigma_v$.
The $\sigma_{\rm int}$ posterior in the $H_0$ inference then simply returns this prior, so the unchanging $\sigma_{\rm int}$ across the variants is in part imposed by this prior rather than being an independent confirmation of the stellar calibration.
The selection edge falls within the observed magnitude range rather than beyond its faint end, so the incompleteness roll-off is sampled by the data: $\mlim=24.07\pm0.19$ mag and $\sigma_{\rm sel}=0.93\pm0.06$ mag in the baseline run.
The full posterior corner of the baseline run is shown in~\cref{fig:representative_corner} and the additional nuisance-parameter posteriors for the main variants in~\cref{tab:trgb_parameter_posteriors}, both in Appendix~\ref{sec:model_priors}.
The mock validation of the pipeline is presented in Appendix~\ref{sec:mock_validation}.

\subsubsection{Posterior predictive checks}\label{sec:results_ppc}

\begin{figure}
\centering
\includegraphics[width=\columnwidth]{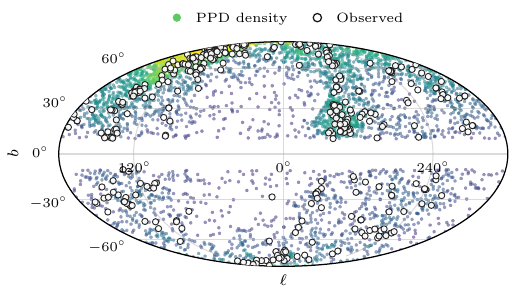}
\caption{Posterior predictive check of the angular host distribution in Galactic coordinates for the Student-$t$ \Manticore\ model with the $48$-pixel angular sky-exposure term.
White circles mark the observed \ac{EDD} hosts, and the coloured points the posterior-predictive sample~(\acs{PPD}), shaded by relative angular density.
The empty band at $|b|<10^\circ$ is the Zone-of-Avoidance mask.
The exposure-weighted forward model reproduces the observed host concentration across both Galactic hemispheres.}
\label{fig:ppc_sky}
\end{figure}

\begin{figure*}
\centering
\includegraphics[width=\textwidth]{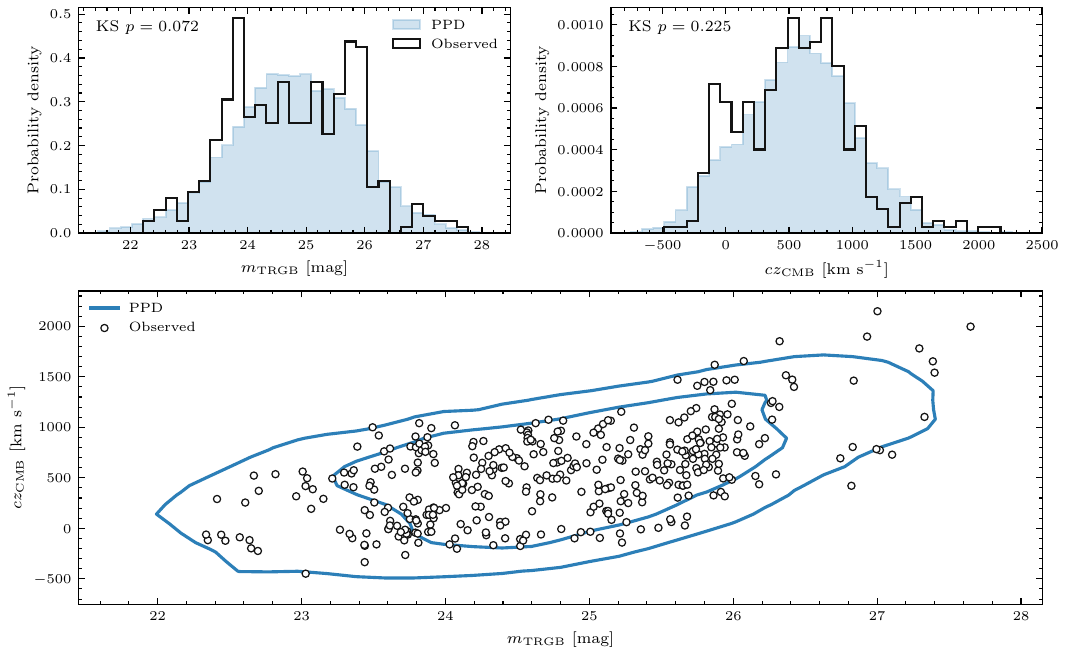}
\caption{Posterior predictive check of the \ac{TRGB} tip-magnitude and \ac{CMB}-frame redshift distributions for the Student-$t$ \Manticore\ model with $4\Mpch$ density smoothing and the $48$-pixel angular sky-exposure term, evaluated at the single dominant \Manticore\ realisation.
Parameters are drawn from the posterior, and mock catalogues forward-modelled through the same selection function and inhomogeneous-Malmquist distance prior used in the inference.
\textit{Top}: marginal $\mobs$ and $c\zCMB$ distributions of the posterior-predictive sample~(\acs{PPD}, blue) and the observed \ac{EDD} hosts (black), each annotated with the two-sample \ac{KS} $p$-value.
\textit{Bottom}: the joint plane, with the $68$ and $95$ per cent posterior-predictive contours and the observed hosts overlaid.
The model reproduces the redshift distribution (\ac{KS} $p=0.23$), while the magnitude comparison is marginal (\ac{KS} $p=0.07$).}
\label{fig:ppc_magnitude_redshift}
\end{figure*}

We check whether the baseline forward model reproduces the observed data through posterior predictive checks, drawing population parameters and $H_0$ from the posterior and forward-modelling mock catalogues through the same selection function and inhomogeneous-Malmquist distance prior used in the inference.
These mock catalogues sample the \ac{PPD}, the distribution of data the inferred model predicts.
We compare their joint tip-magnitude--redshift distribution in~\cref{fig:ppc_magnitude_redshift} and their angular distribution in~\cref{fig:ppc_sky} with the observed \ac{EDD} hosts.
The checks are evaluated on the baseline model (Student-$t$ and the $48$-pixel sky-exposure term) at its single evidence-preferred \Manticore\ realisation.
The forward model reproduces the redshift distribution (two-sample \ac{KS} $p=0.23$), while the magnitude comparison is marginal (\ac{KS} $p=0.07$).
The angular distribution shows no obvious discrepancy~(\cref{fig:ppc_sky}).
The marginal disagreement in the magnitude comparison is driven by two peaks in the observed distribution, near $\mobs\simeq24$ and $\simeq26$ mag, which the smoother \ac{PPD} does not reproduce.

\Cref{fig:ppc_sky} shows the angular check in Galactic coordinates.
The forward model draws host positions from a single \Manticore\ realisation weighted by the source-density model, applies the $|b|\geq10^\circ$ Zone-of-Avoidance mask, and reweights each direction by the inferred angular sky-exposure term of~\cref{sec:angular_selection}.
With the sky-exposure term, the predicted density matches the observed host concentration towards the northern overdensities and the more heavily targeted fields across both Galactic hemispheres.
Without it, the prediction follows the density field alone and over-populates the northern overdensities relative to the data.

\newcommand{\TRGBvariantsbody}{\multicolumn{7}{@{}l}{\textit{Baseline}} \\
{\bfseries\boldmath Student-$t$} & {\boldmath $72.29\pm2.55$} & {\boldmath $57\pm5$} & {\boldmath $0.0$} & {\boldmath $68.4\pm5.9$} & {\boldmath $67\pm7$} & {\boldmath $-1.2\pm2.5$} \\
\addlinespace
\multicolumn{7}{@{}l}{\textit{Redshift likelihood}} \\
Student-$t$ & $72.29\pm2.55$ & $57\pm5$ & $0.0$ & $68.4\pm5.9$ & $67\pm7$ & $-1.2\pm2.5$ \\
Gaussian & $66.97\pm2.65$ & $124\pm5$ & $-36.2$ & $69.4\pm5.5$ & $132\pm7$ & $-32.8\pm4.6$ \\
\addlinespace
\multicolumn{7}{@{}l}{\textit{Density smoothing}} \\
$4\Mpch$, Student-$t$ & $72.29\pm2.55$ & $57\pm5$ & $0.0$ & $68.4\pm5.9$ & $67\pm7$ & $-1.2\pm2.5$ \\
$4\Mpch$, Gaussian & $66.97\pm2.65$ & $124\pm5$ & $-36.2$ & $69.4\pm5.5$ & $132\pm7$ & $-32.8\pm4.6$ \\
$8\Mpch$, Student-$t$ & $72.42\pm2.69$ & $57\pm5$ & $-1.5$ & $68.5\pm5.8$ & $67\pm7$ & $-3.1\pm2.9$ \\
$8\Mpch$, Gaussian & $67.09\pm2.84$ & $124\pm5$ & $-40.2$ & $69.6\pm5.4$ & $132\pm7$ & $-34.8\pm4.7$ \\
No smoothing, Student-$t$ & $64.81\pm2.67$ & $67\pm5$ & $-5.6$ & $68.1\pm6.0$ & $67\pm7$ & $-7.4\pm5.7$ \\
No smoothing, Gaussian & $63.53\pm2.69$ & $130\pm5$ & $-33.1$ & $68.9\pm5.7$ & $132\pm7$ & $-39.1\pm7.0$ \\
\addlinespace
\multicolumn{7}{@{}l}{\textit{Angular sky exposure}} \\
$48$-pixel ($N_{\rm side}=2$), Student-$t$ & $72.29\pm2.55$ & $57\pm5$ & $0.0$ & $68.4\pm5.9$ & $67\pm7$ & $-1.2\pm2.5$ \\
$48$-pixel ($N_{\rm side}=2$), Gaussian & $66.97\pm2.65$ & $124\pm5$ & $-36.2$ & $69.4\pm5.5$ & $132\pm7$ & $-32.8\pm4.6$ \\
$12$-pixel ($N_{\rm side}=1$), Student-$t$ & $72.06\pm2.81$ & $57\pm5$ & $-17.8$ & $68.3\pm5.9$ & $67\pm8$ & $-20.6\pm2.9$ \\
$12$-pixel ($N_{\rm side}=1$), Gaussian & $66.72\pm2.77$ & $124\pm5$ & $-56.5$ & $69.2\pm5.5$ & $132\pm7$ & $-52.3\pm4.9$ \\
No sky exposure, Student-$t$ & $72.15\pm2.70$ & $57\pm5$ & $-45.9$ & $68.4\pm5.9$ & $67\pm8$ & $-45.6\pm3.6$ \\
No sky exposure, Gaussian & $66.83\pm2.92$ & $124\pm5$ & $-81.6$ & $69.4\pm5.6$ & $132\pm7$ & $-77.1\pm5.5$ \\
\addlinespace
\multicolumn{7}{@{}l}{\textit{Flow sector}} \\
Velocity monopole $V_{\rm mono}$, Student-$t$ & $73.63\pm3.05$ & $56\pm5$ & $-0.5$ & $74.7\pm5.2$ & $64\pm8$ & $0.0$ \\
Velocity monopole $V_{\rm mono}$, Gaussian & $69.94\pm3.34$ & $124\pm5$ & $-37.2$ & $74.0\pm5.3$ & $131\pm7$ & $-32.7\pm3.9$ \\
No reconstruction, free $\Vext$, Student-$t$ & $70.40\pm2.94$ & $95\pm7$ & $-167.1$ & -- & -- & -- \\
No reconstruction, free $\Vext$, Gaussian & $70.12\pm3.18$ & $143\pm6$ & $-183.1$ & -- & -- & -- \\
}
\begin{table*}
\centering
\footnotesize
\setlength{\tabcolsep}{3pt}
\begin{tabularx}{\textwidth}{Xcccccc}
\toprule
 & \multicolumn{3}{c}{Evidence-weighted} & \multicolumn{3}{c}{Equal-weight stacked} \\
\cmidrule(lr){2-4}\cmidrule(lr){5-7}
Variant & $H_0$ & $\sigma_v$ & $\Delta\log Z_{\rm ens}$ & $H_0$ & $\sigma_v$ & $\Delta\log Z_{\rm field}$ \\
 & $[\kmsecMpc]$ & $[\kmsec]$ & & $[\kmsecMpc]$ & $[\kmsec]$ & \\
\midrule
\multicolumn{7}{@{}l}{\textit{Baseline}} \\
{\bfseries\boldmath Student-$t$} & {\boldmath $72.29\pm2.55$} & {\boldmath $57\pm5$} & {\boldmath $0.0$} & {\boldmath $68.4\pm5.9$} & {\boldmath $67\pm7$} & {\boldmath $-1.2\pm2.5$} \\
\addlinespace
\multicolumn{7}{@{}l}{\textit{Redshift likelihood}} \\
Student-$t$ & $72.29\pm2.55$ & $57\pm5$ & $0.0$ & $68.4\pm5.9$ & $67\pm7$ & $-1.2\pm2.5$ \\
Gaussian & $66.97\pm2.65$ & $124\pm5$ & $-36.2$ & $69.4\pm5.5$ & $132\pm7$ & $-32.8\pm4.6$ \\
\addlinespace
\multicolumn{7}{@{}l}{\textit{Density smoothing}} \\
$4\Mpch$, Student-$t$ & $72.29\pm2.55$ & $57\pm5$ & $0.0$ & $68.4\pm5.9$ & $67\pm7$ & $-1.2\pm2.5$ \\
$4\Mpch$, Gaussian & $66.97\pm2.65$ & $124\pm5$ & $-36.2$ & $69.4\pm5.5$ & $132\pm7$ & $-32.8\pm4.6$ \\
$8\Mpch$, Student-$t$ & $72.42\pm2.69$ & $57\pm5$ & $-1.5$ & $68.5\pm5.8$ & $67\pm7$ & $-3.1\pm2.9$ \\
$8\Mpch$, Gaussian & $67.09\pm2.84$ & $124\pm5$ & $-40.2$ & $69.6\pm5.4$ & $132\pm7$ & $-34.8\pm4.7$ \\
No smoothing, Student-$t$ & $64.81\pm2.67$ & $67\pm5$ & $-5.6$ & $68.1\pm6.0$ & $67\pm7$ & $-7.4\pm5.7$ \\
No smoothing, Gaussian & $63.53\pm2.69$ & $130\pm5$ & $-33.1$ & $68.9\pm5.7$ & $132\pm7$ & $-39.1\pm7.0$ \\
\addlinespace
\multicolumn{7}{@{}l}{\textit{Angular sky exposure}} \\
$48$-pixel ($N_{\rm side}=2$), Student-$t$ & $72.29\pm2.55$ & $57\pm5$ & $0.0$ & $68.4\pm5.9$ & $67\pm7$ & $-1.2\pm2.5$ \\
$48$-pixel ($N_{\rm side}=2$), Gaussian & $66.97\pm2.65$ & $124\pm5$ & $-36.2$ & $69.4\pm5.5$ & $132\pm7$ & $-32.8\pm4.6$ \\
$12$-pixel ($N_{\rm side}=1$), Student-$t$ & $72.06\pm2.81$ & $57\pm5$ & $-17.8$ & $68.3\pm5.9$ & $67\pm8$ & $-20.6\pm2.9$ \\
$12$-pixel ($N_{\rm side}=1$), Gaussian & $66.72\pm2.77$ & $124\pm5$ & $-56.5$ & $69.2\pm5.5$ & $132\pm7$ & $-52.3\pm4.9$ \\
No sky exposure, Student-$t$ & $72.15\pm2.70$ & $57\pm5$ & $-45.9$ & $68.4\pm5.9$ & $67\pm8$ & $-45.6\pm3.6$ \\
No sky exposure, Gaussian & $66.83\pm2.92$ & $124\pm5$ & $-81.6$ & $69.4\pm5.6$ & $132\pm7$ & $-77.1\pm5.5$ \\
\addlinespace
\multicolumn{7}{@{}l}{\textit{Flow sector}} \\
Velocity monopole $V_{\rm mono}$, Student-$t$ & $73.63\pm3.05$ & $56\pm5$ & $-0.5$ & $74.7\pm5.2$ & $64\pm8$ & $0.0$ \\
Velocity monopole $V_{\rm mono}$, Gaussian & $69.94\pm3.34$ & $124\pm5$ & $-37.2$ & $74.0\pm5.3$ & $131\pm7$ & $-32.7\pm3.9$ \\
No reconstruction, free $\Vext$, Student-$t$ & $70.40\pm2.94$ & $95\pm7$ & $-167.1$ & -- & -- & -- \\
No reconstruction, free $\Vext$, Gaussian & $70.12\pm3.18$ & $143\pm6$ & $-183.1$ & -- & -- & -- \\

\bottomrule
\end{tabularx}
\caption{Posterior medians and standard deviations of $H_0$ and the residual-velocity scale $\sigma_v$, with the Bayesian evidence, for the \ac{EDD} \ac{TRGB} run set under both weighting treatments, grouped by the modelling choice varied relative to the baseline (bold): the \Manticore\ density field smoothed on a $4\Mpch$ scale, the Student-$t$ redshift likelihood, and the $48$-pixel angular sky-exposure term.
Every row but the no-reconstruction pair uses \Manticore.
Every variant is run separately on each of the $80$ realisations, and the two treatments combine those posteriors with evidence weights and with uniform weights.
Because the evidence-weighted posterior typically collapses onto a single realisation, only the equal-weight stack includes the field-to-field scatter, so the two uncertainties are not comparable.
Nor is $\sigma_v$ comparable between the Student-$t$ and Gaussian rows, which parametrise differently shaped residuals.
$\Delta\log Z_{\rm ens}$ and $\Delta\log Z_{\rm field}$ are defined in~\cref{eq:dlogz_ens,eq:dlogz_field}, each quoted relative to the highest-scoring row by that statistic, and $\Delta\log Z_{\rm field}$ with its scatter across the $80$ realisations (\cref{eq:dlogz_field_std}).
A larger $\Delta\log Z$ favours the row in question.
Dashes mark the no-reconstruction variant, which has no realisations to weight.
}
\label{tab:trgb_h0_variants}
\end{table*}

\subsection{Redshift likelihood}\label{ssec:results_zlike}
Holding the density-field smoothing fixed at $4\Mpch$, we test the sensitivity of the inference to the redshift likelihood choice.
The evidence favours the Student-$t$ over the Gaussian in every variant of~\cref{tab:trgb_h0_variants}.
In the baseline configuration the preference is $\Delta\logZ_{\rm field}=+31.6\pm4.4$ and $\Delta\logZ_{\rm ens}=+36.2$.

The two likelihood choices yield different $H_0$ values under evidence weighting: the Gaussian \Manticore\ run gives $H_0=67.0\pm2.6\kmsecMpc$ and the Student-$t$ run $72.3\pm2.6\kmsecMpc$.
Equal-weight stacked, the two instead lie close together, at $69.4\pm5.5$ and $68.4\pm5.9\kmsecMpc$.
The difference in the evidence-weighted $H_0$ arises because the two likelihoods select different dominant realisations, which have different $H_0$ medians.
Field by field, the Student-$t$ likelihood gives an $H_0$ lower than the Gaussian by $0.8\pm0.9\kmsecMpc$.
We therefore report both treatments throughout.

\begin{figure}
    \centering
    \includegraphics[width=\columnwidth]{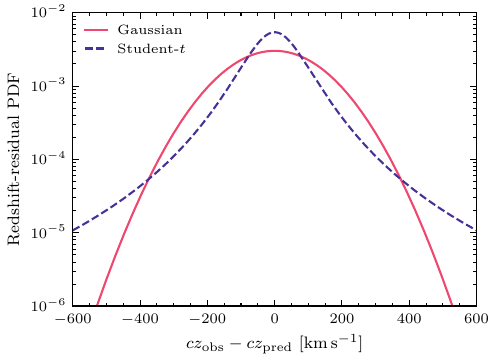}
    \caption{Comparison of the redshift likelihoods of the baseline \Manticore\ run: the Gaussian (solid) and the Student-$t$ (dashed), at their equal-weight stacked inferred scales $\sigma_v=132$ and $67\kmsec$ with the Student-$t$ tail index $\nu=2.59$, on a logarithmic density axis.
    The Student-$t$ model is decisively preferred by the evidence comparison~(\cref{tab:trgb_h0_variants}).}
    \label{fig:trgb_cz_likelihood_pdfs}
\end{figure}

The inferred residual distribution is strongly non-Gaussian, with a tail index of $\nu=2.59\pm0.48$.
At this tail index the distribution is far from Gaussian, so the scale $\sigma_v$ is no longer representative of a physical velocity dispersion.
\Cref{fig:trgb_cz_likelihood_pdfs} contrasts the two residual distributions at their inferred scales, $\sigma_v\simeq67\kmsec$ under the Student-$t$ against $132\kmsec$ under the Gaussian.
These heavy tails likely represent small-scale infall: many \ac{TRGB} hosts cluster around Virgo, Fornax, and Centaurus\,A, where the \texttt{COLA} reconstruction cannot resolve the virial motions.

\subsection{Density smoothing scale}\label{ssec:results_smoothing}
We now adopt the Student-$t$ redshift likelihood and vary the density smoothing scale, retaining the $4\Mpch$ Student-$t$ configuration as the reference.
Under the equal-weight stack, doubling the density smoothing from $4$ to $8\Mpch$ moves the Student-$t$ inference from $H_0=68.4\pm5.9$ to $68.5\pm5.8\kmsecMpc$.
Neither scale is significantly preferred, with the finer $4\Mpch$ leading $8\Mpch$ by $\Delta\logZ_{\rm field}=+1.9\pm2.0$~(\cref{tab:trgb_h0_variants}).
We verify that removing the density smoothing entirely leaves the equal-weight stacked Student-$t$ $H_0$ at $68.1\pm6.0\kmsecMpc$, consistent with the $4$ and $8\Mpch$ values.
The unsmoothed model is nevertheless disfavoured by $\Delta\logZ_{\rm field}=6.2\pm5.9$ relative to $4\Mpch$.

Evidence-weighted, the $4$ and $8\Mpch$ runs likewise agree, at $H_0=72.3\pm2.6$ and $72.4\pm2.7\kmsecMpc$, because the same realisation dominates both.
Removing the smoothing instead moves the evidence-weighted value to $64.8\pm2.7\kmsecMpc$ by changing which realisation dominates the mixture.
At fixed realisation the smoothing scale has little effect: relative to $4\Mpch$, the $8\Mpch$ run shifts $H_0$ by $+0.11\pm0.13\kmsecMpc$ and the unsmoothed run by $-0.26\pm0.43\kmsecMpc$.

\subsection{Angular sky-exposure model}\label{ssec:results_skyexp}

The angular sky-exposure selection term is decisively preferred, at $\Delta\logZ_{\rm field}=+44.4\pm3.9$ and $\Delta\logZ_{\rm ens}=+45.9$ under the Student-$t$ likelihood.
It nevertheless leaves $H_0$ unchanged under both treatments~(\cref{tab:trgb_h0_variants}).

The sky-exposure term is, however, required for the posterior predictive distributions to match the data, as shown in~\cref{fig:ppc_sky} and~\cref{fig:ppc_magnitude_redshift}.
At the dominant realisation the two-sample \ac{KS} $p$-value for the redshift distribution rises from $0.012$ without it to $0.23$ with it, whereas the magnitude $p$-value stays near $0.07$--$0.11$.
We have verified that repeating the posterior predictive check over the $80$ realisations leaves both the redshift and the magnitude $p$-values in the same range as at the highest-evidence realisation, so the qualitative outcome is unchanged.
Inferring the $48$ angular weights is a substitute for what should ideally be done: constructing the sky-exposure map from the actual \ac{HST} pointing footprints of the \ac{EDD} hosts, which the archival compilation does not record.

We test both the $12$-pixel ($N_{\rm side}=1$) and $48$-pixel ($N_{\rm side}=2$) partitions of~\cref{sec:angular_selection}, and the baseline adopts the finer $48$-pixel map.
The evidence favours the $48$-pixel map over the $12$-pixel one by $\Delta\logZ_{\rm field}\simeq+19$ under the Student-$t$ likelihood, while the central value of $H_0$ is unchanged~(\cref{tab:trgb_h0_variants}).
We do not refine the partition further: the $48$-pixel map already has only ${\sim}8$ of the $394$ hosts per pixel on average.
An $N_{\rm side}=4$ partition would leave ${\sim}2$, below the Dirichlet concentration $\alpha=4$ of~\cref{sec:angular_selection}, so the exposure weights would be prior-dominated.
Field by field, the angular term barely moves the inference: at fixed realisation, adding the $48$-pixel term to the no-exposure model shifts $H_0$ by $-0.02\pm0.15\kmsecMpc$, below $0.6\kmsecMpc$ for every one of the $80$ realisations.
Refining from $12$ to $48$ pixels shifts it by $+0.11\pm0.10\kmsecMpc$.

\subsection{Flow modelling}\label{ssec:results_flow}

We now consider how the inferred $H_0$ depends on modifications to the \Manticore\ velocity field.
Unlike the redshift-likelihood, smoothing, and sky-exposure choices examined so far, which shift the equal-weight stacked and field-by-field $H_0$ only slightly, modifications to the velocity model shift the central value of $H_0$ substantially at fixed realisation.
Here we compare the baseline with two modifications, adding a velocity monopole and removing the reconstruction, both in the Student-$t$, $48$-pixel configuration (the Gaussian likelihood gives similar results, as shown in~\cref{tab:trgb_h0_variants}).
Adding a velocity monopole to the baseline configuration raises the equal-weight stacked $H_0$, while the evidence-weighted value moves by only $+1.3\kmsecMpc$.
Evidence-weighted, it gives $V_{\rm mono}=-9.7\pm10.5\kmsec$ and $H_0=73.6\pm3.0\kmsecMpc$.
Equal-weight stacked, it gives $V_{\rm mono}=-38.5\pm27.2\kmsec$ and $H_0=74.7\pm5.2\kmsecMpc$ under the Student-$t$ likelihood.
The evidence neither favours nor disfavours the inclusion of the monopole, at $\Delta\logZ_{\rm field}=+1.2\pm2.5$ and $\Delta\logZ_{\rm ens}=-0.5$ relative to the baseline.
Field by field, adding the monopole raises $H_0$ by $5.4\pm3.4\kmsecMpc$ relative to the baseline, with $94$ per cent of the $80$ realisations moving upward.
The dominant realisation is atypical in this respect, its monopole shift being only $+1.4\kmsecMpc$, which is why the evidence-weighted value moves so much less than the equal-weight one.
Finally, a no-reconstruction Student-$t$ control with $\Vext$ inferred and no velocity monopole gives $H_0=70.4\pm2.9\kmsecMpc$ but is decisively disfavoured, at $\Delta\logZ_{\rm ens}=-167.1$.
This control has no \Manticore\ realisation to marginalise over and therefore no field-to-field spread.
Its $|\Vext|=562\pm12\kmsec$ absorbs the entire local flow.

At the observer position the \Manticore\ reconstruction predicts a velocity of $368\pm45\kmsec$ towards $(\ell,\,b)=(233\pm14,\,47\pm11)^\circ$, where the uncertainties are the scatter across the $80$ realisations.
The external flow of the baseline model is $|\Vext|=337\pm8\kmsec$ towards $(\ell,\,b)=(285\pm2,\,-3\pm1)^\circ$ evidence-weighted, and $388\pm43\kmsec$ towards $(296\pm5,\,-1\pm8)^\circ$ equal-weight stacked, the latter including the field-by-field spread.
Adding the two per realisation yields a Local-Group-frame dipole of $616\pm29\kmsec$ towards $(\ell,\,b)=(272\pm4,\,27\pm3)^\circ$ equal-weight stacked, and $657\pm9\kmsec$ towards $(269\pm1,\,28\pm1)^\circ$ evidence-weighted.
The equal-weight stacked value sits $0.1\sigma$ in amplitude and $2.7^\circ$ in direction from the \textit{Planck} \ac{CMB}-frame Local-Group value of $620\pm15\kmsec$ towards $(271.9\pm2.0,\,29.6\pm1.4)^\circ$~\citep{Planck_2020_overview}.
The dipole is more tightly constrained than either term entering it, because across the $80$ realisations the reconstructed velocity at the observer and the inferred $\Vext$ are strongly anti-correlated, with Pearson coefficients of $-0.89$, $-0.73$, and $-0.93$ in the three Cartesian components.
The anti-correlation arises because $\Vext$ absorbs whatever coherent motion the reconstruction misses locally.
The inferred $|\Vext|$ is a factor of ${\sim}3$ larger than the $111\pm15\kmsec$ that we obtain by calibrating the \Manticore\ reconstruction against the Cosmicflows-4 Tully--Fisher catalogue~\citep{Kourkchi_2020A}, whose distances we infer from the $W1$-band calibration, following the procedure of~\citet{VF_olympics}.
The two need not describe the same motion.
Against a deep sample, $\Vext$ is forced to describe the coherent motion of the entire reconstructed volume, whereas over the shallow \ac{TRGB} volume ($cz\lesssim2150\kmsec$) it is instead free to absorb misspecification of the reconstructed flow locally.
Its amplitude here should therefore not be read as a bulk motion of the entire box.
The closer comparison is instead with the extragalactic Cepheid hosts, where the same \Manticore\ reconstruction and the same $\Vext$ parametrisation give $|\Vext|=258\pm66\kmsec$~\citep{Stiskalek_2026,Stiskalek_2026_erratum}, $1.2\sigma$ from our evidence-weighted baseline and $1.7\sigma$ from the equal-weight one.

\section{Discussion}\label{sec:discussion}

We first discuss the local density and velocity field, which sets the limiting uncertainty of this measurement~(\cref{ssec:disc_flow}), then the selection, galaxy-bias, and calibration systematics that are subdominant to it~(\cref{ssec:disc_systematics}), and finally the comparison with previous work~(\cref{ssec:disc_literature}).
Our central argument is that the precision of this two-rung ladder is set by the reconstruction of the local flow, and that the stellar and selection systematics become relevant only once that limit is lifted.

\subsection{The local density and velocity field}\label{ssec:disc_flow}

The limiting factor in this inference is not the \ac{TRGB} calibration but our knowledge of the local density and velocity field that the \ac{TRGB} hosts probe.
The evidence-weighted $H_0$ uncertainty is $2.6\kmsecMpc$, whereas across the $80$ \Manticore\ realisations the field-median $H_0$ scatters by $5.2\kmsecMpc$ in the baseline inference.
That scatter is comparable to the difference between the \ac{SHOES} and \textit{Planck} \acs{LCDM} values that a local measurement is meant to arbitrate.
The hosts lie at $c\zCMB\lesssim2150\kmsec$ and are concentrated far below that limit, with $87$ per cent of them within $1000\kmsec$ and half within $560\kmsec$.
While the constrained simulations suppress cosmic variance on the largest scales, such a small volume leaves $H_0$ sensitive to structures at the limit of what \Manticore\ can reconstruct.

The evidence-weighted $H_0$ is determined by a single dominant \Manticore\ field~(\cref{ssec:results_baseline}).
The marginalisation is formally well motivated, but here its outcome is set by whichever realisation happens to dominate.
The equal-weight stack is not an alternative marginalisation but a measure of the reconstruction-to-reconstruction uncertainty, so we quote the two side by side.
Most of the variant-to-variant shifts in the evidence-weighted $H_0$ follow from this.
At fixed realisation the redshift likelihood, density smoothing, and sky-exposure choices each move $H_0$ by less than $1\kmsecMpc$, yet across those same variants the evidence-weighted value spans $63.5$ to $72.4\kmsecMpc$, because the dominant realisation changes between them.
The velocity monopole is the sole exception: it raises $H_0$ by $5.4\pm3.4\kmsecMpc$ at fixed realisation but by only $1.3\kmsecMpc$ after evidence weighting.
The evidence picks out the \Manticore\ Student-$t$ class as the favoured model and strongly disfavours every Gaussian-likelihood variant relative to that baseline.

We take the density and velocity fields from \Manticore~\citep{McAlpine_2025} and treat them as given.
That makes the inference conditional on that set of posterior samples of the local Universe, on the \TWOMPP\ data and \acs{LCDM} prior that constrain it, and on the other \ac{BORG} modelling assumptions that generate it, mainly the reconstruction resolution, the galaxy-bias model, the count likelihood, and the particle-mesh gravity solver.
The clearest sign that these assumptions leave the reconstructed flow incomplete is internal to the reconstruction: at the observer position \Manticore\ predicts $368\pm45\kmsec$ where the Local-Group dipole requires $620\pm15\kmsec$, a deficit of ${\sim}250\kmsec$ that the ``external'' bulk flow has to model.
Added to the reconstructed velocity at the observer, the equal-weight stacked $\Vext$ recovers the \textit{Planck} Local-Group dipole in both amplitude and direction~(\cref{ssec:results_flow}).

Under the Gaussian likelihood we infer $\sigma_v=124\pm5\kmsec$ evidence-weighted and $132\pm7\kmsec$ equal-weight stacked~(\cref{tab:trgb_h0_variants}).
These sit $1.8\sigma$ and $1.5\sigma$ respectively from the $173\pm27\kmsec$ obtained for the extragalactic Cepheid hosts of~\citet{Stiskalek_2026} under the same likelihood and the same \Manticore\ reconstruction.
That redshift likelihood class is nonetheless the one the evidence rejects, by $\Delta\logZ_{\rm field}=31.6\pm4.4$, so the physically meaningful residual is the one implied by the baseline Student-$t$.
For a scale $\sigma_v$ and tail index $\nu$ the implied root mean square is $\sigma_v\sqrt{\nu/(\nu-2)}$, which for the equal-weight stacked $(67\pm7\kmsec,\,2.59\pm0.48)$ gives $141\kmsec$, and for the evidence-weighted $(57\pm5\kmsec,\,2.34\pm0.34)$ gives $149\kmsec$.
The incompleteness of the reconstruction over this volume therefore shows up mainly in the coherent $\Vext$, not in an inflated scatter relative to the Cepheid hosts.

This large $\Vext$ is not wholly unexpected, because the reconstructed flow at the observer position is colder than the \ac{CMB} dipole would require~\citep{McAlpine_2025}.
The reconstruction is constrained by the \TWOMPP\ redshift survey, whereas the dipole can be sourced in part by structure beyond that volume~\citep{Peacock_1992,Lavaux_2010,Bilicki_2011,Stiskalek_2026B}.
We nevertheless adopt \Manticore\ over the linear-theory reconstruction of~\citet{Carrick_2015} and other literature fields~\citep{Lilow_2021,Courtois_2023} because it is favoured by the Bayesian evidence when evaluated against direct distance tracers~\citep{McAlpine_2025,VF_olympics}.
It also supplies an ensemble of posterior samples rather than a single point estimate, which is what lets us quantify the residual cosmic variance at all.

Two data improvements would reduce this sensitivity to the flow model.
The first is a \ac{TRGB} sample extending well beyond the present limit, where the peculiar velocity is a few per cent of the recession velocity rather than the tens of per cent it reaches here.
At those depths the residual sensitivity to non-linear peculiar velocities is small: relative to a \TWOMPP-based reconstruction, ignoring them raises $H_0$ by only ${\sim}\,0.4\pm0.5\kmsecMpc$ over $0.023<z<0.046$~\citep{Tsaprazi_2025}.
The \ac{JWST} has already measured a tip at $30\pm4\Mpc$~\citep{Carleton_2024}, and~\citet{Anand_2021} project \ac{TRGB} distances out to $30$--$40\Mpc$ with modest exposure times, so the required depth is within reach.
That first measurement is a favourable case rather than a typical host: a quiescent low-surface-brightness dwarf with minimal crowding and $94$ stars surviving the photometric cuts.
The exposure required for a crowded disc at the same distance is therefore longer.
What such programmes do not supply is a sample suited to a two-rung ladder: their targets are chosen because they host \acp{SN}~Ia or have distances from surface-brightness fluctuations~\citep{Blakeslee_2023,Li_2024a,Freedman_2025}, so the selection is inherited mainly from the third rung rather than specified in terms of the tip observables.

The second is a constraint on the local initial conditions from data more local than the \TWOMPP\ redshift survey.
Conditioning the reconstruction on such data would mean drawing realisations that explain the specific volume the \ac{TRGB} hosts occupy, which would be expected to narrow the field-to-field spread.
This would require extending the \ac{BORG} framework to direct distance tracers, a step demonstrated on mock data by~\citet{PrideauxGhee_2023}.
Bayesian reconstruction from peculiar velocities alone has already been applied to real catalogues~\citep{Boruah_2022}.
Alternative reconstructions of the local density and velocity field from direct distance tracers already exist~\citep{Courtois_2023,Valade_2022,Valade_2026}, but none of them targets the very local volume we consider here.
There is nonetheless strong precedent for pushing \ac{BORG} to even smaller scales: the positions and peculiar velocities of nearby galaxies have been used to constrain the Milky Way--Andromeda pair itself~\citep{Wempe_2024,Wempe_2026}.
We do not use those reconstructions here because they do not explicitly model the Virgo Cluster, the dominant structure and source of infall within the \ac{TRGB} volume.

A spatially constant radial inflow or outflow, the velocity monopole $V_{\rm mono}$, is unphysical over a large volume and partly degenerate with $H_0$ because a uniform radial flow mimics a change in the recession rate.
The degeneracy is not perfect, because the recession velocity grows linearly with distance whereas $V_{\rm mono}$, as we formulate it, does not, so a sample spanning a range of distances separates the two.
Over the shallow \ac{TRGB} volume the monopole is nonetheless a defensible first-order nuisance term, absorbing any residual monopole in the reconstruction itself.
We therefore carry it as a variant rather than in the baseline.
The inferred $V_{\rm mono}=-38.5\pm27.2\kmsec$ is consistent with zero, so the variant provides no evidence for a net radial flow within the \ac{TRGB} volume.
Constraining the local initial conditions jointly with the more local data discussed above would allow any residual monopole in the reconstruction to be modelled explicitly, rather than absorbed into a free nuisance term.

\subsection{Selection, galaxy bias, and calibration}\label{ssec:disc_systematics}

When an archival compilation has no record of how its entries were assembled, its selection function can only be posited and partially constrained by the data in retrospect.
This makes the inference conditional on a model that may be misspecified.
We posit a selection on the tip apparent magnitude.
Any additional selection would have to act on a different observable, most plausibly the redshift, or the magnitude selection itself would have to depart from the simple cumulative-normal form we assume.

We do not include a redshift-dependent selection term, even though velocity plays a part in how the parent sample was assembled:~\citet{Karachentsev_2013} include a galaxy if $V_{\rm LG}<600\kmsec$ \emph{or} $D<11\Mpc$ (\cref{ssec:sel_model}).
Because either condition alone suffices, a galaxy above the velocity threshold still enters the list on its distance.
Redshift is in any case a poor distance indicator in this volume, since peculiar velocities of ${\sim}300\kmsec$ and Virgo infall of ${\sim}180\kmsec$ are comparable to the $600\kmsec$ threshold itself.
Since distance is also not an observable, we posit an effective selection on the apparent tip magnitude instead.
The magnitude selection we posit reflects the depth of the imaging:~\citet{Anand_2021} measure the tip out to just past $20\Mpc$, a limit set by the exposure time needed to resolve the red giant branch rather than by a sharp failure of the method.
The posterior predictive checks of~\cref{fig:ppc_magnitude_redshift} support a magnitude-only selection, in that the magnitude-based model reproduces the observed redshift distribution of the hosts.
No redshift-dependent selection is therefore required to explain the data, although the magnitude comparison itself remains marginal.

A further question that neither the selection term nor that check probes is how the hosts trace the underlying density field.
The dominant \ac{HST} programme targets known unobscured galaxies within $10\Mpc$, the range over which a single orbit resolves the red giant branch well enough for a tip measurement~\citep{Anand_2021}.
We model this galaxy population with the inhomogeneous-Malmquist prior, whose bias term we can only posit.
We verify that the equal-weight stacked $H_0$ moves by less than $1\kmsecMpc$ across the variants that shift the inferred bias slope from $\alpha_{\rm low}=2.25\pm0.39$ to $4.26\pm0.59$~(\cref{tab:trgb_h0_variants,tab:trgb_parameter_posteriors}), so this dependence is subdominant to the field-to-field scatter.
We infer the limiting magnitude to be $\mlim=24.07\pm0.19$ mag with a smooth transition width $\sigma_{\rm sel}=0.93\pm0.06$ mag.
The evidence prefers the angular sky-exposure term at $\Delta\logZ_{\rm field}=+44.4\pm3.9$, plausibly because the archival campaigns targeted rich regions of the sky rather than sampling the volume uniformly.
Surveys that fix their filters, depth, and sky coverage in advance would ideally make these parameters known rather than inferred.
The proposed \textit{Roman} wide-area programme would do so~\citep{Blakeslee_2023}, its selection well specified even though its targets are inherited from the surface-brightness-fluctuation and \ac{SN} rungs, as would ground-based resolved-star searches whose completeness is calibrated against image simulations~\citep{MutluPakdil_2021}.

A further astrophysical uncertainty is the assumed galaxy-bias model, which maps the \Manticore\ matter field to the expected host number density entering the inhomogeneous-Malmquist term.
This mapping is required in any full physical forward model, but is poorly constrained.
At the level of dark matter haloes the bias is a power law in the local density with an exponential downturn that suppresses occupation in voids~\citep{Neyrinck_2014}.
\Manticore\ generalises this to a sigmoid-truncated double power law, adding a second slope at high density and replacing the exponential cutoff with a sigmoid~\citep{McAlpine_2025}.
The form we adopt in~\cref{eq:bias_manticore}, here and in our previous Cepheid-only analysis~\citep{Stiskalek_2026}, retains the two power-law regimes but not the low-density truncation.
\citet{Bartlett_2024} show, however, that local-in-density bias models do not reproduce the clustering of the halo field on the smoothing scales and at the halo masses relevant to our $4\Mpch$ field.
Moreover, the \ac{TRGB} hosts are predominantly nearby low-mass galaxies, a regime in which the bias prescriptions we adopt have never been calibrated.
\citet{Neyrinck_2014}, for example, characterise the bias of dark matter haloes rather than of the galaxies occupying them.
The mapping we require is therefore harder than the halo case that already fails, since it must additionally describe the occupation of low-mass galaxies within those haloes.

This matters most for a sample as nearby as ours, whose distance uncertainties localise individual hosts within specific structures, so the bias model acts on resolved features of the density field rather than on a smooth average.
As a related example, when comparing the second data release of the Zwicky Transient Facility (ZTF DR2) Type~Ia \acp{SN} with the same \Manticore\ reconstruction,~\citet{GillesLordet_2026} find that the \ac{SN} rate is not a linear tracer of the matter density, reaching local enhancements of a factor of two to five within nearby clusters.
\Cref{tab:trgb_parameter_posteriors} and~\cref{fig:representative_corner} give the posteriors of $\bm{b}$ across the variants.

The \ac{TRGB} calibration contributes two further systematics.
The tip is standardised at a fixed colour pivot with the slope held at $\alpha_c=0.2$~\citep{Anand_2022}, so any residual colour or metallicity dependence propagates into $M_0$.
The \ac{EDD} compilation also aggregates measurements from heterogeneous \ac{HST} programmes, where a different edge-detection algorithm can shift the inferred tip~\citep{Scolnic_2023}.
The two enter the inference differently.
The colour-pivot systematic acts coherently across the sample, so it enters as a single offset in $M_0$ and hence a single offset in $H_0$.
The algorithmic heterogeneity instead varies between programmes and enters as additional scatter in the individual tip magnitudes.
Neither systematic can generate the $5.2\kmsecMpc$ field-to-field spread in $H_0$, because the same tip magnitudes enter the inference for every field.
A coherent offset $\Delta M_0$ shifts $H_0$ by approximately the same fraction, ${\approx}\,0.46\,\Delta M_0$, in every field, so it moves the headline value but not the spread between fields.
The programme-to-programme scatter is absorbed by $\sigma_{\rm int}$, whose prior is inferred from the same heterogeneous tip magnitudes and so already includes it.

\subsection{Comparison with previous work}\label{ssec:disc_literature}

Our constraint is consistent with the published \ac{TRGB} ladders that retain the \ac{SN} third rung, but is less precise than the \ac{CCHP} and \texttt{CATS} values by a factor of ${\sim}1.2$--$1.4$ under the evidence-weighted treatment and ${\sim}3$ under the equal-weight one.
The consistency therefore follows in large part from that lower precision.
The \ac{CCHP} infers $H_0=70.4\pm1.2\,({\rm stat})\pm1.3\,({\rm sys})\pm0.7\,(\sigma_{\rm SN})\kmsecMpc$~\citep{Freedman_2025,Freedman_2025_erratum}, the \texttt{CATS} reanalysis gives $73.2\pm2.1\kmsecMpc$~\citep{Scolnic_2023}, and a ladder built exclusively from \acp{SN}~Ia in quiescent, early-type hosts gives $75.3\pm2.9\kmsecMpc$~\citep{Newman_2026}.
That calibrator sample is deliberately disjoint from the star-forming hosts on which Cepheid calibrations rely.
\citet{Scolnic_2023} decompose the $3.4\kmsecMpc$ separating their value from the earlier \ac{CCHP} result of $69.8\pm1.9\kmsecMpc$~\citep{Freedman_2019} into $2.0\kmsecMpc$ from \ac{SN}-survey and peculiar-flow corrections absent in that work and $1.4\kmsecMpc$ from the inhomogeneity of the \ac{TRGB} calibration across the ladder.
\citet{Freedman_2023} challenge two of the tip identifications that enter that reanalysis.
For NGC\,4038 and NGC\,4536, the tip measurements that \texttt{CATS} weights most highly give distance moduli $0.7$ and $0.6$ mag smaller, respectively, than the \ac{SHOES} Cepheid moduli for the same hosts.
The offsets are attributed to the unsupervised edge detection selecting the tip of the asymptotic giant branch rather than the \ac{TRGB}, with NGC\,4038 as the explicit example.
Those analyses use the \ac{TRGB} only to calibrate a second rung reaching $cz\gtrsim3000\kmsec$, where peculiar velocities are a few per cent of the recession velocity.

The \ac{TRGB}-only measurements that precede ours confront the same flow dependence, but address it by modelling infall onto a single structure.
\citet{Kim_2020} obtain $65.8\pm3.5\,({\rm stat})\pm2.4\,({\rm sys})\kmsecMpc$ from galaxies falling towards Virgo, solving a spherical infall model that includes $\Omega_\Lambda$ and inferring $H_0$ jointly with the Virgo turnaround radius $R_0=6.76\pm0.35\Mpc$ and a residual velocity scatter $\sigma_v=62\pm14\kmsec$.
The same infall approach was applied separately to the Local Group and to Virgo by~\citet{Peirani_2006}, and later extended to Coma with distances other than the \ac{TRGB}~\citep{Benisty_2026}.
We instead predict the flow at every host position from the reconstructed field, which replaces a single-structure infall model with a more complete description of the local density and velocity field.

\section{Conclusions}\label{sec:conclusions}

Both high-precision local anchors of the Hubble tension rely on the Type~Ia \ac{SN} rung~\citep{Riess_2022,Freedman_2025}.
In this paper we have tested how well the local ladder constrains $H_0$ without that rung, using only the \ac{TRGB} stellar indicator and the host redshifts and geometric anchors, following the Cepheid-only precedent of~\citet{Stiskalek_2026}.
We have forward-modelled the tip magnitudes and their latent colours (standardised at the fixed slope $\alpha_c=0.2$), the host redshifts, the magnitude-limited catalogue selection with its Galactic-plane mask and angular sky-exposure variants, and the \ac{LMC} and NGC\,4258 geometric anchors.
We conditioned the inference on each of the $80$ \Manticore\ realisations of the local density and peculiar-velocity fields in turn.
We then stacked the resulting posteriors, weighted either equally or by their Bayesian evidence.
Our main findings are as follows.

\begin{enumerate}
    \item[(i)] The limiting uncertainty is the local density and especially velocity field, rather than the \ac{TRGB} calibration.
    The field-median $H_0$ scatters by $5.2\kmsecMpc$ across the $80$ \Manticore\ realisations, comparable to the \ac{SHOES}--\textit{Planck}-\acs{LCDM} difference.
    \item[(ii)] The equal-weight stack of the $80$ single-field posteriors gives $H_0=68.4\pm5.9\kmsecMpc$, whereas weighting the same posteriors by their evidence gives $72.3\pm2.6\kmsecMpc$ in the baseline inference.
    The baseline adopts a Student-$t$ redshift likelihood and is the highest-evidence variant tested.
    Interpreting the per-field evidences as relative model probabilities gives an effective sample size of $N_{\rm eff}=1.00$ out of $80$, so this constraint has collapsed onto a single \Manticore\ realisation.
    The equal-weight stack is more conservative and follows from treating all realisations as equally plausible.
    The evidence-weighted value effectively assumes the single highest-evidence realisation to be true, which is likely sensitive to the particular sampling of the \Manticore\ \ac{BORG} inference.
    \item[(iii)] At fixed realisation the redshift likelihood, density smoothing, and sky-exposure choices each move $H_0$ by less than $1\kmsecMpc$.
    Only the addition of a velocity monopole moves it appreciably, raising $H_0$ by $5.4\pm3.4\kmsecMpc$ at fixed realisation.
\end{enumerate}

This supernova-free distance ladder therefore does not yet deliver a precise $H_0$.
Across the variants of the velocity model, the equal-weight stacked $H_0$ ranges from $68.4\pm5.9$ to $74.7\pm5.2\kmsecMpc$, so a \ac{TRGB}-only ladder does not have the constraining power to distinguish the \textit{Planck} and \ac{SHOES} values.
The main limitation is the small volume the \ac{TRGB} hosts probe.
The density field enters through the inhomogeneous-Malmquist distance prior, whereas the velocity at the host positions maps directly into $H_0$ through the redshift residuals.
A \ac{TRGB}-only ladder is far more sensitive to this than the geometry and Cepheid ladder of~\citet{Stiskalek_2026} because $87$ per cent of the \ac{TRGB} hosts lie at $c\zCMB<1000\kmsec$, where the flow is a large fraction of the recession velocity.
This limitation is therefore not intrinsic to a two-rung ladder.
Our Cepheid-only analysis uses the \ac{SHOES} Cepheid hosts, which reach $3300\kmsec$~\citep{Riess_2022}.
There the same framework and the same \Manticore\ reconstruction give $H_0=71.1\pm1.4\kmsecMpc$, with much weaker sensitivity to the density- and velocity-field treatment~\citep{Stiskalek_2026,Stiskalek_2026_erratum}.

The archival nature of the sample is a further limitation: the \ac{EDD} compilation has no clear record of how its entries were assembled, so the magnitude and angular selection terms must be posited and inferred rather than specified a priori.
The same is true of our Cepheid analysis~\citep{Stiskalek_2026,Stiskalek_2026_erratum}.
In the \ac{TRGB}-only ladder the uncertainty from this unspecified selection is likely subdominant to the residual cosmic variance of the local flow.
It nevertheless motivates recording the selection of future \ac{TRGB} samples, so that the selection function can be specified rather than inferred.

More constructively, the collapse of the evidence-weighted posterior onto a single realisation shows how sensitive this inference is to the assumed model of the local Universe.
The remaining gains are therefore in the modelling of the local Universe itself, in the density field that enters the inhomogeneous-Malmquist distance prior and especially in the velocity field that maps into the redshift residuals.
At these distances $H_0$ cannot be inferred without modelling the peculiar velocity of every host.
Narrowing the field-to-field spread requires reconstructions constrained by data within the volume the hosts occupy, or a \ac{TRGB} sample deep enough that the peculiar velocity is a smaller fraction of the recession velocity so that existing reconstructions such as \Manticore\ are sufficient.
This matters beyond the present measurement, because the local volume anchors the calibration of the \acp{SN}: how it is modelled propagates into every $H_0$ inferred through that rung.

\section{Data Availability}

The \ac{EDD} \ac{TRGB} catalogue is described by~\citet{Anand_2021}.
The \Manticore\ realisations are available at \url{https://www.cosmictwin.org}.
The code underlying this article will be shared on reasonable request to the corresponding author.

\section*{Acknowledgements}

We thank Wendy Freedman, Alan Heavens, Barry Madore, and Adam Riess for helpful discussions and feedback on this work.
We also thank Jonathan Patterson for smoothly running the Glamdring Cluster hosted by the University of Oxford, where part of the data processing was performed.
This work used the University of Oxford Advanced Research Computing facility (\url{https://doi.org/10.5281/zenodo.22558}).
RS is supported by a Hintze Fellowship at the Oxford Centre for Astrophysical Surveys, funded through generous support from the Hintze Family Charitable Foundation.
RS acknowledges financial support from STFC Grant No.\ ST/X508664/1 and the Snell Exhibition of Balliol College, Oxford.
HD is supported by a Royal Society University Research Fellowship (grant no.\ 211046).
GL acknowledges support from the Simons Foundation through the Simons Collaboration on ``Learning the Universe'' (SFI-MPS-LU-00008515-03).
This work was carried out within the Aquila Consortium (\url{https://aquila-consortium.org}).

\bibliographystyle{mnras}
\bibliography{ref}

\clearpage
\appendix
\onecolumn

\clearpage

\section{Model parameters: priors and posteriors}\label{sec:model_priors}

\begin{figure*}
\centering
\includegraphics[width=\textwidth]{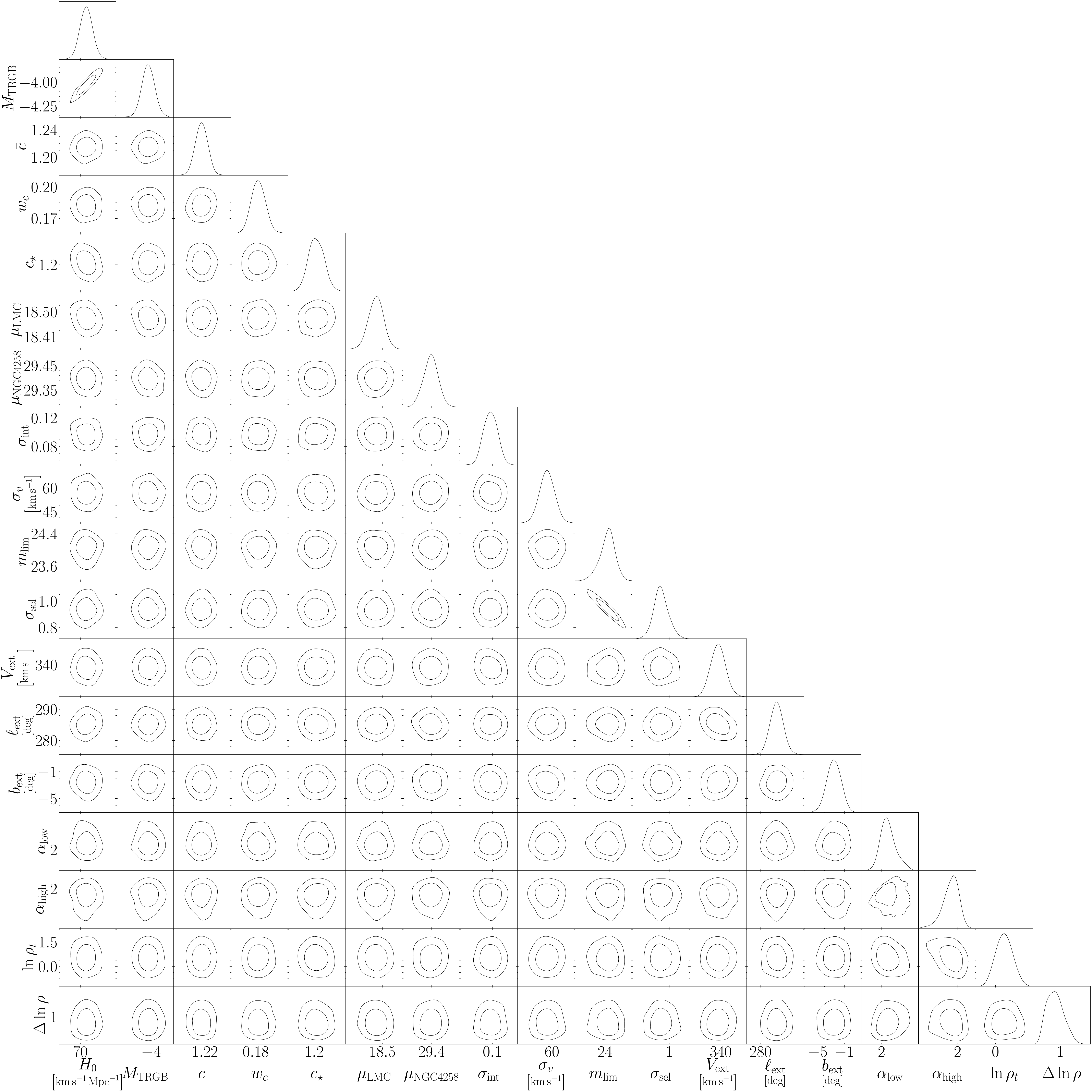}
\caption{Posterior corner for the baseline \Manticore\ run with $4\Mpch$ density smoothing: marginal one-dimensional densities and two-dimensional credible contours for $H_0$, the \ac{TRGB} calibration, the anchor moduli, the selection edge, $\Vext$, the residual $\sigma_v$, and the \Manticore\ source-density parameters.
Contours denote the $1\sigma$ and $2\sigma$ credible regions.
}
\label{fig:representative_corner}
\end{figure*}

\Cref{tab:trgb_priors} lists the priors on the sampled global and nuisance parameters,~\cref{tab:trgb_parameter_posteriors} their marginal posteriors across the main variants, and~\cref{fig:representative_corner} the full joint posterior of the baseline run.

\begin{center}
\centering
\captionsetup{type=table,hypcap=false}
{\footnotesize
\begin{tabularx}{\textwidth}{lXp{0.15\textwidth}p{0.24\textwidth}}
\toprule
Parameter & Prior & Run & Role \\
\midrule
\multicolumn{4}{@{}l}{\textbf{\textit{Cosmology}}} \\
$H_0$ & $\mathcal{U}(40,\,100)\kmsecMpc$ & All & Hubble constant \\
\addlinespace
\midrule
\multicolumn{4}{@{}l}{\textbf{\textit{\acs{TRGB} calibration and anchors}}} \\
$M_0$ & $\mathcal{N}(-4.05,\,0.5^2)$ mag & All & \ac{TRGB} absolute magnitude at colour $c_\star$ \\
$\alpha_c$ & $\delta(0.2)$ mag/mag & All & Colour--magnitude slope of $M_{\rm F814W}^{\rm TRGB}$ at the pivot $c_\star$, fixed to the \ac{EDD} value \\
$c_\star$ & $\mathcal{N}(1.23,\,0.1^2)$ mag & All & F606W--F814W colour pivot \\
$\sigma_{\rm int}$ & $\mathcal{N}(0.1,\,0.01^2)$ mag, $\sigma_{\rm int}>0.01$ mag & All & Residual \ac{TRGB} scatter \\
$\bar c$ & $\mathcal{U}(0,\,3)$ mag & All & Mean of the latent \ac{EDD} colour-standardisation parent \\
$w_c$ & $\mathcal{U}(0.001,\,1)$ mag & All & Width of the latent \ac{EDD} colour-standardisation parent \\
$\mu_{\rm LMC}$ & $p(\mu)\propto10^{3\mu/5}$ on $17<\mu<20$ mag & All & \ac{LMC} anchor distance modulus \\
$\mu_{\rm N4258}$ & $p(\mu)\propto10^{3\mu/5}$ on $28<\mu<31$ mag & All & NGC\,4258 anchor distance modulus \\
\addlinespace
\midrule
\multicolumn{4}{@{}l}{\textbf{\textit{Velocity field and redshift likelihood}}} \\
$\sigma_v$ & $\mathrm{Maxwell}(\mu=300\kmsec)$ & All & Residual velocity dispersion \\
$\nu$ & $\mathcal{N}(30,\,10^2)$, $1<\nu<100$ & Student-$t$ redshift variants & Redshift-likelihood degrees of freedom \\
$|\Vext|$ & $\mathcal{U}(0,\,1000)\kmsec$ & All & External bulk-flow amplitude \\
$\phi_{V_{\rm ext}},\,\cos\theta_{V_{\rm ext}}$ & $\phi_{V_{\rm ext}}\hookleftarrow\mathcal{U}(0,\,2\pi)$, $\cos\theta_{V_{\rm ext}}\hookleftarrow\mathcal{U}(-1,\,1)$ & All & External bulk-flow direction \\
$V_{\rm mono}$ & $\mathcal{U}(-500,\,500)\kmsec$ & Velocity-monopole variants & Constant radial velocity monopole added to the reconstructed field \\
\addlinespace
\midrule
\multicolumn{4}{@{}l}{\textbf{\textit{Source density}}} \\
$\alpha_{\rm low}$ & $\mathcal{N}(1.5,\,1^2)$, $\alpha_{\rm low}>0$ & All & Low-density galaxy-bias slope \\
$\alpha_{\rm high}$ & $f\,\alpha_{\rm low}$, $f\hookleftarrow\mathcal{N}(0.5,\,0.5^2)$, $0<f<1$ & All & High-density galaxy-bias slope \\
$\ln\rho_t$ & $\mathcal{N}(0.5,\,0.5^2)$ & All & Galaxy-bias transition density \\
$\Delta_{\ln\rho}$ & $\mathcal{N}(0.5,\,0.5^2)$, $0.05\le\Delta_{\ln\rho}\le3$ & All & Galaxy-bias transition width \\
\addlinespace
\midrule
\multicolumn{4}{@{}l}{\textbf{\textit{Selection}}} \\
$\mlim$ & $\mathcal{U}(22.101,\,29)$ mag & All & \ac{TRGB}-magnitude selection threshold \\
$\sigma_{\rm sel}$ & $\mathcal{U}(0.15,\,1.5)$ mag & All & \ac{TRGB}-magnitude selection width \\
$b_{\rm min}$ & $\delta(10^\circ)$ & All & Galactic-plane selection mask \\
$\bm{\theta}$ & $\mathrm{Dirichlet}(\alpha\bm{1})$, $\alpha=4$ fixed ($\kappa=48,\,192$) & Sky-exposure variants & Angular exposure weights on $N_{\rm pix}=12,\,48$ \texttt{HEALPix} pixels \\
\bottomrule
\end{tabularx}
}
\captionof{table}{Priors on sampled global and nuisance parameters.
Columns give the parameter, its prior, the runs in which it enters, and its role in the model, with ``All'' denoting a parameter present in every run.
$\delta(\cdot)$ denotes a parameter held fixed at the quoted value.
Normal priors are quoted as mean and variance, and an accompanying inequality gives the range to which the prior is truncated.
The Maxwell prior is quoted by its mean $\mu=2a\sqrt{2/\pi}$ rather than its scale $a$.
The adopted $\mu=300\kmsec$ corresponds to $a=188\kmsec$ and a mode $\sqrt{2}\,a=266\kmsec$.
Informative priors are the truncated-normal prior on $\sigma_{\rm int}$ obtained from the tip magnitudes alone (\cref{ssec:results_baseline}), and the Maxwell prior on $\sigma_v$, whose mean is set near the expected linear-theory \ac{LOS} peculiar-velocity dispersion.
The \Manticore\ double-power-law slopes are sampled subject to $\alpha_{\rm high}\le\alpha_{\rm low}$, so the high-density slope does not exceed the low-density slope.
The $c_\star$ prior is centred on the~\citet{Anand_2022} pivot value.}
\label{tab:trgb_priors}
\end{center}

\begin{table*}
\centering
\scriptsize
\setlength{\tabcolsep}{3pt}
\renewcommand{\arraystretch}{1.15}
\begin{tabular}{lcccccccc}
\toprule
Variant & Smoothing & $c_\star$ & $\mu_{\rm LMC}$ & $\mu_{\rm N4258}$ & Selection & Density parameters & $\nu$ & External flow \\
 & $[\Mpch]$ & $[\rm mag]$ & $[\rm mag]$ & $[\rm mag]$
 & \begin{tabular}[t]{@{}c@{}}$m_{\rm lim}$,\\$\sigma_{\rm sel}$\end{tabular}
 & \begin{tabular}[t]{@{}c@{}}$\alpha_{\rm low}$,\\$\alpha_{\rm high}$,\\$\ln\rho_t$,\\$\Delta_{\ln\rho}$\end{tabular} & &
 \begin{tabular}[t]{@{}c@{}}$|\Vext|~[\kmsec]$,\\$\ell~[^\circ]$,\\$b~[^\circ]$,\\$V_{\rm mono}~[\kmsec]$\end{tabular} \\
\midrule
\multicolumn{9}{@{}l}{\textbf{\textit{Redshift likelihood} ($48$-pixel sky exposure)}} \\
Gaussian & $4$ & $1.22\pm0.10$ & $18.48\pm0.03$ & $29.40\pm0.03$
 & \begin{tabular}[t]{@{}c@{}}$23.97\pm0.18$,\\$0.92\pm0.05$\end{tabular}
 & \begin{tabular}[t]{@{}c@{}}$2.33\pm0.39$,\\$1.73\pm0.42$,\\$0.55\pm0.51$,\\$0.72\pm0.39$\end{tabular}
 & --
 & \begin{tabular}[t]{@{}c@{}}$355\pm11$,\\$293\pm3$,\\$-5\pm2$,\\--\end{tabular} \\
\addlinespace
Student-$t$ (baseline) & $4$ & $1.23\pm0.10$ & $18.48\pm0.03$ & $29.40\pm0.03$
 & \begin{tabular}[t]{@{}c@{}}$24.07\pm0.19$,\\$0.93\pm0.06$\end{tabular}
 & \begin{tabular}[t]{@{}c@{}}$2.27\pm0.38$,\\$1.71\pm0.42$,\\$0.53\pm0.52$,\\$0.75\pm0.40$\end{tabular}
 & $2.34\pm0.34$
 & \begin{tabular}[t]{@{}c@{}}$337\pm8$,\\$285\pm2$,\\$-3\pm1$,\\--\end{tabular} \\
\midrule
\multicolumn{9}{@{}l}{\textbf{\textit{Density smoothing} (Student-$t$)}} \\
$8\Mpch$ smoothing & $8$ & $1.22\pm0.10$ & $18.48\pm0.03$ & $29.40\pm0.03$
 & \begin{tabular}[t]{@{}c@{}}$24.07\pm0.19$,\\$0.95\pm0.06$\end{tabular}
 & \begin{tabular}[t]{@{}c@{}}$4.25\pm0.59$,\\$3.31\pm1.11$,\\$0.80\pm0.50$,\\$0.50\pm0.39$\end{tabular}
 & $2.35\pm0.35$
 & \begin{tabular}[t]{@{}c@{}}$337\pm8$,\\$285\pm2$,\\$-3\pm1$,\\--\end{tabular} \\
\midrule
\multicolumn{9}{@{}l}{\textbf{\textit{Angular sky exposure} (Student-$t$)}} \\
No sky exposure & $4$ & $1.23\pm0.10$ & $18.48\pm0.03$ & $29.40\pm0.03$
 & \begin{tabular}[t]{@{}c@{}}$24.09\pm0.18$,\\$0.92\pm0.05$\end{tabular}
 & \begin{tabular}[t]{@{}c@{}}$2.51\pm0.39$,\\$1.87\pm0.43$,\\$0.60\pm0.54$,\\$0.69\pm0.42$\end{tabular}
 & $2.33\pm0.35$
 & \begin{tabular}[t]{@{}c@{}}$337\pm8$,\\$285\pm2$,\\$-3\pm1$,\\--\end{tabular} \\
\addlinespace
$12$-pixel ($N_{\rm side}=1$) & $4$ & $1.23\pm0.10$ & $18.48\pm0.03$ & $29.40\pm0.03$
 & \begin{tabular}[t]{@{}c@{}}$24.06\pm0.18$,\\$0.94\pm0.05$\end{tabular}
 & \begin{tabular}[t]{@{}c@{}}$2.18\pm0.39$,\\$1.62\pm0.43$,\\$0.57\pm0.58$,\\$0.72\pm0.41$\end{tabular}
 & $2.35\pm0.36$
 & \begin{tabular}[t]{@{}c@{}}$337\pm8$,\\$285\pm2$,\\$-3\pm1$,\\--\end{tabular} \\
\midrule
\multicolumn{9}{@{}l}{\textbf{\textit{Flow sector} (Student-$t$, $48$-pixel sky exposure)}} \\
Velocity monopole & $4$ & $1.23\pm0.10$ & $18.48\pm0.03$ & $29.40\pm0.03$
 & \begin{tabular}[t]{@{}c@{}}$24.05\pm0.19$,\\$0.94\pm0.06$\end{tabular}
 & \begin{tabular}[t]{@{}c@{}}$2.26\pm0.44$,\\$1.67\pm0.41$,\\$0.52\pm0.52$,\\$0.73\pm0.40$\end{tabular}
 & $2.31\pm0.35$
 & \begin{tabular}[t]{@{}c@{}}$338\pm9$,\\$285\pm2$,\\$-2\pm1$,\\$-10\pm10$\end{tabular} \\
\addlinespace
No reconstruction & -- & $1.23\pm0.10$ & $18.48\pm0.03$ & $29.40\pm0.03$
 & \begin{tabular}[t]{@{}c@{}}$24.12\pm0.19$,\\$0.94\pm0.05$\end{tabular}
 & -- & $3.50\pm0.75$
 & \begin{tabular}[t]{@{}c@{}}$562\pm12$,\\$287\pm2$,\\$29\pm1$,\\--\end{tabular} \\
\bottomrule
\end{tabular}
\caption{Additional posterior summaries for the \ac{EDD} \ac{TRGB} model variants in~\cref{tab:trgb_h0_variants}, under the evidence-weighted treatment.
Entries are posterior medians with $1\sigma$ standard deviations, given line by line in the order listed in the column header.
A dash denotes a parameter not sampled in that variant.
Rows are grouped by the modelling choice varied relative to the baseline, the Student-$t$ row.
The Gaussian counterparts of the density-smoothing and sky-exposure groups shift the tabulated nuisance parameters by less than one standard deviation.
The velocity monopole $V_{\rm mono}$ is sampled only in the monopole variant, and the no-reconstruction variant carries no density field and hence no density parameters.
Because $N_{\rm eff}=1$, every \Manticore\ row is the posterior of that variant's highest-evidence realisation.
}
\label{tab:trgb_parameter_posteriors}
\end{table*}

\section{Mock validation}\label{sec:mock_validation}

We validate the forward model with mock \ac{TRGB} catalogues.
The validation checks the recovery of injected parameters when the assumed data-generating process is true.
It therefore tests the implementation of the hierarchical model and of the assumed selection formulation, rather than the adequacy of the selection model for the real \ac{EDD} sample.

For each mock catalogue we draw host positions uniformly within a sphere around the observer and retain them with probability proportional to the source-density model, so the accepted hosts follow the inhomogeneous-Malmquist distance prior $\tilde{\pi}(r)\propto r^2\,n(\bm{r},\,\delta(\bm{x}),\,\bm{b})$ of~\cref{eq:distance_prior}.
For reconstructed mocks the density and velocity fields are evaluated from a single \Manticore\ realisation, with the same galaxy-bias model in the mock generation and in the recovery.
We deposit this field using piecewise-cubic-spline mass assignment, smooth the density on a $4\Mpch$ scale, leave the velocity field unsmoothed, impose the $|b|\geq10^\circ$ mask, and omit the angular sky-exposure term.
We use the priors of the baseline configuration of the paper, with the recovery model omitting the angular sky-exposure term as in the mocks.
Using the same field in generation and recovery removes model misspecification from the test, so any residual bias reflects the implementation alone.
Generating and analysing the mocks with different fields would additionally probe sensitivity to that misspecification, which we do not test here.

The accepted latent host distance and sky position correspond to a cosmological redshift and a \ac{LOS} peculiar velocity, which combine into the ``observed'' redshift.
We draw the latent dereddened colour from a Gaussian parent population and add a measurement error to form the observed colour.
We then generate the observed \ac{TRGB} magnitude from the latent colour as $M_0 + \alpha_c(c-c_\star) + \mu$, with the intrinsic \ac{TRGB} scatter $\sigma_{\rm int}$ and F814W measurement noise added in quadrature.
The \ac{LMC} and NGC\,4258 anchors are drawn from the same geometric-distance and tip-magnitude uncertainties used in the inference.

\begin{figure}
\centering
\includegraphics[width=0.5\columnwidth]{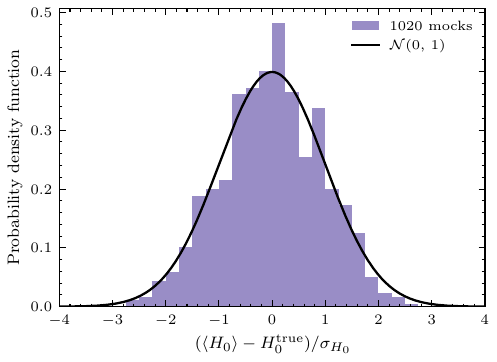}
\caption{
    Distribution of the $H_0$ standardised bias, $(\langle H_0\rangle-H_0^{\rm true})/\sigma_{H_0}$, across the $1020$ mock catalogues (purple), against the standard normal $\mathcal{N}(0,\,1)$ (black).
Here $\langle H_0\rangle$ and $\sigma_{H_0}$ are the posterior mean and standard deviation of each mock and $H_0^{\rm true}=72.2\kmsecMpc$.
The $H_0$ posteriors are well calibrated across the mock suite.}
\label{fig:trgb_mock_h0_bias}
\end{figure}

Catalogue membership is imposed by the same soft \ac{TRGB}-magnitude window used in the model.
The lower magnitude cut is fixed at the injected value, while the limiting magnitude and transition width are inferred as nuisance parameters in the recovery run.
Each mock is analysed with the standard inference pipeline.
We repeat this over $1020$ mock realisations and record the posterior standardised bias of $H_0$ as a calibration diagnostic.
A correctly specified model should give a mean standardised bias consistent with zero and scatter close to unity.
\Cref{fig:trgb_mock_h0_bias} shows the distribution of the $H_0$ standardised bias across the $1020$ mocks.
A \ac{KS} test does not distinguish the distribution from the standard normal ($p=0.73$), so the $H_0$ posteriors are well calibrated across the suite.

\bsp
\label{lastpage}
\end{document}